\documentclass[12pt]{article}
\pdfoutput=1
\usepackage{amsmath, amsfonts, amsthm, amssymb,}
\usepackage{epsfig}
\usepackage{slashed}
\usepackage{epstopdf}
\usepackage{latexsym}
\usepackage{graphicx}
\usepackage{xspace}
\usepackage{rotating}
\usepackage{booktabs}
\usepackage{enumitem}
\usepackage{xspace}
\usepackage[numbers,compress]{natbib}
\usepackage{multirow}
\usepackage{multicol}
\usepackage{bigstrut}

\usepackage[T1]{fontenc}

\usepackage[margin=20pt,small]{caption}
\usepackage{subcaption}

\usepackage[utf8]{inputenc}
\usepackage[babel]{csquotes}
\usepackage[toc]{appendix}

\usepackage{color}
\usepackage{datetime}
\usepackage[
colorlinks=false,
linkcolor=darkblue,  
urlcolor=blue,    
filecolor=blue,     
citecolor=red,
linktocpage=true,
pdfstartview=FitV,
bookmarksopen=true,
hidelinks
]{hyperref}

\ifpdf
\fi

\newcommand*{\boxedcolor}{black}
\makeatletter
\renewcommand{\boxed}[1]{\textcolor{\boxedcolor}{%
		\fbox{\normalcolor\m@th$\displaystyle#1$}}}
\makeatother

\definecolor{cardinal}{rgb}{0.6,0,0}
\definecolor{darkgreen}{rgb}{0,0.5,0}
\definecolor{golden}{rgb}{0.92, 0.7, 0}
\definecolor{midnight}{rgb}{0, 0, 0.5}
\definecolor{darkblue}{rgb}{0.2, 0, 0.8}

\newcommand{\pf}{\mathfrak{p}}

\newcommand{\OO}{\mathcal O}

   \let\d=\delta

     \let\y=\psi
  \let\D=\Delta

\newtheorem{theorem}{Theorem}

\newcommand{\id}{{\mathbf 1}}

\newcommand{\op}{\ensuremath{\mathcal{O}}\xspace}

\newcommand{\vev}[1]{\ensuremath{\langle #1 \rangle}\xspace}

\newcommand{\du}[2]{_{ #1 }^{\phantom{ #1 } #2 }}

\newcommand{\bxx}[1]{\begin{#1}}
\newcommand{\be}{\bxx{equation}}
\newcommand{\ee}{\end{equation}}

\def\ea{\end{array}}

\def\sl{\frak{sl}}

\newcommand{\gbullet}{{\xspace\bullet\xspace}}
\newcommand{\wbullet}{{\xspace\circ\xspace}}

\begin{document}  
	
	\begin{titlepage}
	\flushright{DESY 20-082}
		
		\medskip
		\begin{center} 
			{\Large \bf Line and surface defects for the free scalar field}

			\bigskip
			\bigskip
			\bigskip
			
			{\bf Edoardo Lauria$^{1}$, Pedro Liendo$^2$, Balt C. van Rees$^{1}$ and Xiang Zhao$^{1}$\\ }
			\bigskip
			\bigskip
			${}^{1}$
			CPHT, CNRS, Institut Polytechnique de Paris, France\\
			\vskip 5mm
			${}^{2}$
			DESY Hamburg, Theory Group, \\
			Notkestraße 85, D-22607 Hamburg, Germany
			\vskip 5mm
			
			\texttt{edoardo.lauria@polytechnique.edu,~pedro.liendo@desy.de,\\
				~balt.van-rees@polytechnique.edu,~xiang.zhao@polytechnique.edu;} \\
		\end{center}
		
		\bigskip
		\bigskip
		
		\begin{abstract}
			\noindent For a single free scalar field in $d \geq 2$ dimensions, almost all the unitary conformal defects must be `trivial' in the sense that they cannot hold interesting dynamics. The only possible exceptions are monodromy defects in $d \geq 4$ and co-dimension three defects in $d \geq 5$. As an intermediate result we show that the $n$-point correlation functions of a conformal theory with a generalized free spectrum must be those of the generalized free theory.
		\end{abstract}
		
		\noindent

	\end{titlepage}
	

\setcounter{tocdepth}{2}

\tableofcontents
\newpage

\section{Introduction and summary}
\label{sec:intro}
Defects are useful probes in quantum field theories. For a given field theory there are often infinitely many different defects and universal results are hard to come by. This situation improves if we focus on long distances where we recover the \emph{conformal defects} which correspond to the (forced) symmetry breaking pattern
\be \label{symmetrybreaking}
\mathfrak{so}(d+1,1) \quad \longrightarrow \quad \mathfrak{so}(p+1,1) \times \mathfrak{so}(d-p)\,.
\ee
The rationale for this pattern is as follows. First, if we assume locality and reflection positivity then a $d$-dimensional infrared theory generally has $\mathfrak{so}(d+1,1)$ conformal invariance. If we now put the $p$-dimensional defect on an $\mathbb{R}^p$ subspace of $\mathbb{R}^d$ then we can assume that it preserves: (a) a $p$-dimensional Poincar\'e symmetry, (b) rotations in the transverse $d-p$ dimensions, and (c) an overall dilatation symmetry in the infrared. There are exceptions to (a) and (b), see for example \cite{Kobayashi:2018okw} for the kinematics of defects charged under transverse rotations, but we will not consider this here. As for (c), a simple computation involving the bulk stress tensor shows that this scale invariance is enhanced to $\mathfrak{so}(p+1,1)$, so $p$-dimensional conformal invariance, if the defect does not contain a specific `virial current' of dimension $p-1$, see for example \cite{Paulos:2015jfa}. In this precise sense the pattern in \eqref{symmetrybreaking} is considered to be the generic situation at long distances.

In the following we will follow standard notation and introduce
\be
q = d - p
\ee
as the co-dimension of the $p$-dimensional defect. In this work we will consider $q > 1$ and, since $p > 0$, $d > 2$ as well. The case $q=1$ is analyzed in \cite{Behan:2020nsf}.\footnote{See also \cite{Prochazka:2019fah} for recent work on free scalars that  interact through a boundary.}

For the physics of the defect both sides of \eqref{symmetrybreaking} are important. Away from the defect the $\mathfrak{so}(d+1,1)$ symmetry algebra acts on the local bulk operators and implies the existence of a convergent \emph{bulk operator product expansion}. On the defect the local operators are organized in representations of $\mathfrak{so}(p+1,1) \times \mathfrak{so}(q)$, with the first factor acting as the usual conformal algebra in $p$ dimensions and the latter as a usual global symmetry -- although neither of these symmetries is generated by a local current on the defect. Furthermore, these defect operators have their own convergent \emph{defect operator product expansion}. The connection with the bulk operators is provided by the \emph{bulk-defect operator expansion} which states that a local bulk operator in the vicinity of the defect can be written as a sum over defect operators. For example, for a scalar operator and a co-dimension two defect we can write:
\begin{align}\label{dOPEscalargen}
	\phi(\vec{x},z,\bar{z})=\sum_{k}\left(\frac{b_\phi^k \,\,\bar{z}^{s_k}}{|z|^{\Delta_{\phi}-\widehat{\Delta}_{k}+s_{k}}}
	\mathfrak{C}_{\widehat{\Delta}_k}[|z|, \vec \nabla^2] \widehat{\mathcal{O}}_k(\vec{x})+\text{c.c.}\right)
\end{align}
where we split the $d$-dimensional Euclidean coordinates as $x^\mu = (\vec{x}, \text{Re}(z), \text{Im}(z) )$ with the latter two coordinates taken to be orthogonal to the defect. The index $k$ labels the different primary defect operators $\widehat{\mathcal{O}}_k$, which in this case are all scalars and therefore labeled by their scaling dimensions $\widehat{\Delta}_k$ and $SO(2)$ spin $s_k$. The contribution of their descendants is taken into account by the (explicitly known \cite{Billo:2016cpy}) differential operator $\mathfrak{C}_{\widehat{\Delta}_k}[|z|, \vec \nabla^2]$. The expansion also furnishes the bulk-defect operator expansion coefficients $b_\phi^{k}$.

For co-dimension two defects the following comment is in order. In equation \eqref{dOPEscalargen} the spins $s$ are integers if the bulk fields are to be single-valued around the defect. This is however not necessary. If the bulk theory has a global symmetry $G$ then one might alternatively require that
\begin{align}
	\phi(\vec{x}, e^{2\pi i}z,e^{-2\pi i}\bar{z})=\phi^g(\vec{x},z),\, \text{for}\,\, g\in G.
\end{align}
For non-trivial $g$ such defects are called \emph{monodromy defects}. One may think of them as the boundaries of the co-dimension one defects that implement $g$. We will only consider $G = {\mathbb Z}_{2}$ and then there is a single type of monodromy defect corresponding to the non-trivial element of $G$. In the presence of such a defect the odd bulk operators have a bulk-defect expansion of the form given in equation \eqref{dOPEscalargen} with half-integer $s$. For more general $G$, like the case studied in \cite{Soderberg:2017oaa}, the expansion would need further modifications.

The philosophy of the \emph{defect bootstrap} is to explore the consistency conditions that follow from the associativity of the three operator expansions given above. In recent years there has been significant progress on this programme \cite{Liendo:2012hy,Gaiotto:2013nva,Gliozzi:2015qsa,Gliozzi:2016cmg,Liendo:2016ymz,Lemos:2017vnx,Hogervorst:2017kbj,Bissi:2018mcq,Kaviraj:2018tfd,Mazac:2018biw,Liendo:2019jpu}. Just as in the ordinary (bulk) conformal bootstrap, it is essential to know the relevant conformal blocks which group the contributions of an entire conformal representation. Pioneering work in this direction was done by \cite{McAvity:1995zd,McAvity:1993ue} in the case of co-dimension 1, whereas \cite{Billo:2016cpy,Gaiotto:2013nva,Gadde:2016fbj,Lauria:2017wav,Guha:2018snh,Lauria:2018klo,Isachenkov:2018pef} contains results for higher co-dimensions.

Ideally the defect bootstrap would lead to a classification of all the possible defects for a given bulk CFT. In the future it might for example be possible to show that the monodromy defect is the only non-trivial line defect in the three-dimensional Ising model, or that the known co-dimension two and four defects are the only conformal defects in the six-dimensional $(2,0)$ theories. In this paper we consider a more modest problem: that of the classification of defects in the theory of a single real free scalar. Our most important conclusion is that there is very little scope for non-trivial conformal defects of co-dimension two and higher in such theories. We consider this somewhat surprising: for example, we do not expect this conclusion to hold for co-dimension one (boundaries). Indeed, for $d > 2$ several non-trivial boundary conditions appear possible \cite{Dimofte:2017tpi,Giombi:2019enr,Herzog:2017xha,Gupta:2019qlg,Herzog:2018lqz,Behan:2020nsf} and for $d = 2$ there exists a family of conformal boundary conditions for a free (compact) scalar \cite{Gaberdiel:2001zq}. Also, non-trivial defects do exist in other cases where the bulk is free, like the non-trivial co-dimension two monodromy defects for a free hypermultiplet in 4d with $\mathcal N=2$ \cite{Cordova:2017mhb,Bianchi:2019sxz} (see also \cite{Gaiotto:2011tf,Gaiotto:2012xa} for not necessarily conformal defects in this theory) and the co-dimension four surface operators in the abelian $(2,0)$ theory \cite{Ganor_1997,Henningson_1999,henningson1999surface,Gustavsson_2003,Gustavsson_2004,Drukker:2020dcz,Mezei:2018url}. Another example are the infinity of possible boundary conditions in a free four-dimensional Maxwell theory \cite{Witten:2003ya}, see also \cite{Gaiotto:2008ak,Seiberg:2016gmd,DiPietro:2019hqe}, which of course also features conformal Wilson and 't Hooft lines.

\subsection{Summary}
Although this work contains some more general results, our main outcome is that most defects in the free scalar theory are `trivial' in the sense that there is no room for any interesting dynamics on the defect: up to potentially an undetermined one-point function (for $q=p+2$ only), all the $n$-point correlation functions of the bulk field $\phi$ are completely fixed.\footnote{More precisely, the connected two-point function of $\phi$ is simply the unique Klein-Gordon propagator with boundary conditions on the defect defined as below, and the connected higher-point functions of $\phi$ are all zero. This definition relies on the Gaussian properties of the bulk scalar field theory, which are not spoiled by a trivial defect. In the literature the word `trivial defect' is often used to mean `no defect'. This definition of `trivial' agrees with ours only in the case of non-monodromy defects without one-point functions.} More precisely, we will show that, in a reflection positive setup,
\begin{itemize}
	\item monodromy defects with $q = 2$ can be non-trivial only if $d \geq 4$;
	\item non-monodromy defects can be non-trivial only if $q = 3$ and if $d \geq 5$;
\end{itemize}
Our reasoning proceeds as follows.

First, in section \ref{sec:preliminaries} we show that the equation of motion for two-point functions strongly constrains the bulk-defect operator expansion of the bulk field $\phi$. We will discuss how, in all cases except the ones given above, this expansion is \emph{completely fixed}. For example, for co-dimension two it must take the form
\begin{align}\label{dOPEscalarphi}
	\phi(\vec{x},z,\bar{z})=\sum_{s}\left(b_\phi^{+,s} \,\,\bar{z}^{s}
	\mathfrak{C}_{\Delta_\phi + s}[|z|, \vec \nabla^2] \psi^{(+)}_s(\vec{x})+\text{c.c.}\right)\,,
\end{align}
for some operators $\psi^{(+)}_s$ with dimensions $\widehat \Delta_s^{(+)} = \Delta_\phi + s$ and transverse spins $s \geq 0$ constrained to be either half-integer or integer depending on the monodromy type of the defect. The coefficients $b_\phi^{+,s}$ are given below.

An expansion like equation \eqref{dOPEscalargen} completely fixes the two-point function of the bulk field $\phi$, but more work is required to also constrain the higher-point functions: we have to learn about the defect OPE of the operators $\psi^{(+)}_s$ themselves. This we do in sections \ref{sec:constrainingDef} and \ref{sec:theorem}, where we will demonstrate that the operators $\psi^{(+)}_s$ are generalized free fields and their $n$-point functions are given by a sum over Wick contractions. In more detail, in section \ref{sec:constrainingDef} we analyze the singularities in the three-point function of one free bulk and two defect operators. Requiring the absence of unphysical singularities implies that the defect OPE of two $\psi^{(+)}_s$ operators can only contain non-trivial operators of the `double twist' type. This analysis however cannot fix the OPE coefficients nor the multi-OPEs of the $\psi^{(+)}_s$ operators. To finish the proof we therefore need one more ingredient and this is provided in section \ref{sec:theorem}: we can use a dispersion relation in the complex time plane for the $n$-point functions of the $\psi^{(+)}_s$ operators. Since the discontinuities in this dispersion relation are trivial the $n$-point functions must be trivial as well, and so our claim of the triviality of the $n$-point functions of the bulk field $\phi$ also follows.

In section \ref{ss:4ptfunctions} we will specialize to the case of line defects. Although the derivation is \emph{grosso modo} the same, some subtleties arise because the analyticity properties of conformal correlation functions on a line are different. However if we assume parity on the defect then our conclusions remain the same. This in particular rules out a non-trivial parity-preserving monodromy defect for a real scalar in $d = 3$, in sharp contrast with the non-trivial supersymmetric monodromy defects in $d = 4$.

Section \ref{sec:confpertthy} is devoted to perturbative tests of our results. We consider examples in conformal perturbation theory that could lead to a non-trivial defect for the free scalar theory and therefore a counterexample to our main claim. As expected these attempts fail, but they do so in a rather interesting manner.

Some applications of our results will be discussed in section \ref{sec:discussion}.

\section{The two-point function of the free scalar}
\label{sec:preliminaries}
In this section we will analyze the spectrum of operators appearing in the bulk-defect operator expansion \eqref{dOPEscalargen} for a free scalar field $\phi$. To do so it suffices to look at two-point functions involving one bulk field $\phi$ and a defect operator $\widehat{\OO}$. By imposing the equation of motion $\square \phi = 0$ (away from contact points), we will find that the spectrum in the bulk-defect operator expansion is highly constrained. We will then consider the two-point function of $\phi$ to fix almost all the coefficients.
In the main text we will focus on $q = 2$ for simplicity of notation. The case with $q > 2$ is discussed in appendix \ref{app:freedOPE}. 

This section is mostly a review of results that have already appeared in the literature. The defect blocks for the scalar two-point function in the presence of the twist defect were first presented in \cite{Gaiotto:2013nva}. For generic $p,q$ defects, the blocks for the scalar two-point functions were computed in \cite{Billo:2016cpy}  (see also \cite{Lauria:2017wav,Lauria:2018klo}). The constraints imposed by the bulk equation of motion and by unitarity on the bulk-defect expansion of a free scalar were first discussed in \cite{Billo:2013jda} (see also \cite{Gaiotto:2013nva}) for the case of the twist defect. This analysis was extended to generic $p$ and $q$ in Appendix B of \cite{Billo:2016cpy}, and the equation ``$s \leq (4-q)/2$'' of that reference is a less refined version of the information presented in the two tables below. For $q=1$, the blocks for the scalar two-point function were obtained in \cite{McAvity:1995zd} while the spectrum of boundary modes of the free scalar was discussed in \cite{Liendo:2012hy,Dimofte:2012pd,Gaiotto:2014gha,Gliozzi:2015qsa}.

\subsection{General form of the two-point functions}
In this section we consider the two-point function of a general scalar bulk primary $\phi$ and a defect primary operator $\widehat{\OO}$. For a defect operator with transverse spin $s$ and scaling dimension $\widehat{\Delta}_{\widehat{\OO}}$ one finds that\footnote{See \cite{Billo:2016cpy,Gaiotto:2013nva,Gadde:2016fbj,Lauria:2017wav,Guha:2018snh,Lauria:2018klo,Isachenkov:2018pef} for recent work on kinematical contraints for defect CFTs and also \cite{Liendo:2012hy,McAvity:1993ue,McAvity:1995zd} for previous studies in the co-dimension one case.}
\begin{align}\label{phiO2pt}
	\langle \phi(\vec{x},z,\bar{z})\widehat{\mathcal{O}}_{-s}(0) \rangle = \frac{b_\phi^{\widehat{\mathcal{O}}}\,\,\bar{z}^s}{|z|^{\Delta_{\phi}-\widehat{\Delta}_{\widehat{\mathcal{O}}}+s}{(|z|^2+|\vec{x}|^2)^{\widehat{\Delta}_{\widehat{\mathcal{O}}}}}}.
\end{align}
where we used the same conventions for parallel and transverse coordinates as listed in the introduction. Recall that in CFTs without defects the functional form of three-point functions efficiently encapsulates the contribution of descendants in the bulk OPE. In the defect setup the two-point function in equation \eqref{phiO2pt} similarly encodes the contribution of descendants in the bulk-defect operator expansion. This is analyzed in appendix \ref{app:freedOPE}; the corresponding infinite-order differential operator is given in equation \eqref{bOPEphi}.

Next we consider the two-point functions of two general bulk scalar operators $\phi$,
\begin{align}\label{phiphi}
	\langle\phi(\vec{x}_1,z_1,\bar{z}_1)\phi(\vec{x}_2,z_2,\bar{z}_2)\rangle.
\end{align}
This correlation function depends non-trivially on two cross-ratios, which we can take to be:
\begin{align}\label{chideftot}
	{{\chi}}\equiv\frac{|\vec{x}_{12}|^2+{|z_1|}^2+{|z_2|}^2}{{|z_1|}{|z_2|}},\quad \varphi=\arg\left(\frac{{z}_1}{{z}_2} \right).
\end{align}
In the following we will need the \emph{defect channel decomposition} of this two-point function which is obtained by plugging in the bulk-defect operator expansion twice. This leads to two infinite-order differential operators of the form given in equation \eqref{bOPEphi} acting on the two-point function of a defect primary. In appendix \ref{app:3ptfunctions} we resum these contributions from the defect descendants and obtain the \emph{defect channel} decomposition:
\begin{align}\label{bulktodefplus}
	\langle \phi(\vec{x}_1,z_1,\bar{z}_1)\phi(\vec{x}_2,z_2,\bar{z}_2)\rangle=\frac{1}{(|z_1||z_2|)^{\Delta \phi }}\sum_s\sum_{\widehat{\mathcal{O}}}|b_\phi^{\widehat{\mathcal{O}}}|^2
	\,\,{e^{-i\,s\,\varphi}}{}\mathcal{F}_{\widehat{\Delta}_{\widehat{\mathcal{O}}}}({\chi}).
\end{align}
where we introduced the \emph{defect conformal blocks} as:
\begin{align}\label{2ptblocksdef}
	\mathcal{F}_{\widehat{\Delta}_{\widehat{\mathcal{O}}}}({\chi})=\chi^{-\widehat{\Delta}_{\widehat{\mathcal{O}}}}\, _2F_1\left(\frac{\widehat{\Delta}_{\widehat{\mathcal{O}}}}{2},\frac{\widehat{\Delta}_{\widehat{\mathcal{O}}}+1}{2};\widehat{\Delta}_{\widehat{\mathcal{O}}}+1-\frac{p}{2};\frac{4}{\chi^2}\right).
\end{align}
We remark that these functions can also be computed by solving certain Casimir equations with appropriate boundary conditions \cite{Gaiotto:2013nva,Billo:2016cpy,Lauria:2017wav,Lauria:2018klo}. One could also consider a \emph{bulk channel decomposition} of the same two-point function in terms of a sum over bulk one-point functions. We will not need this decomposition in our analysis.

\subsection{Two-point functions of the free scalar}
\label{ss:twopoint}
We now specialise to the case where $\phi$ is a free bulk scalar of canonical dimension $\Delta_\phi=\frac{d}{2}-1$ and therefore obeys $\square \phi = 0$ away from contact points.

For the bulk-defect two-point function given in \eqref{phiO2pt} the action of the Laplacian gives
\begin{align}\label{BDfreeeq}
	0=\langle \square\phi(\vec{x},z,\bar{z})\widehat{\mathcal{O}}_{-s}(0) \rangle \sim (\widehat{\Delta}_{\widehat{\mathcal{O}}}-\Delta_\phi+|s|)(\widehat{\Delta}_{\widehat{\mathcal{O}}}-\Delta_\phi-|s|)\frac{b_\phi^{\widehat{\mathcal{O}}}\,\bar{z}^{s}}{|z|^{2+\Delta_{\phi}-\widehat{\Delta}_{\widehat{\mathcal{O}}}+s}{(|z|^2+|\vec{x}|^2)^{\widehat{\Delta}_{\widehat{\mathcal{O}}}}}}.
\end{align}
Therefore, the only defect primaries allowed to appear in the bulk-to-defect OPE of a free scalar belong to one of the two families that we denote as $\psi_{s}^{(\pf)}$ with $\pf=\pm$, with dimensions $\widehat{\Delta}_{s}^{(\pf)}$ given by:\footnote{Note that $\widehat{\Delta}_{s}^{(+)} + \widehat{\Delta}_{s}^{(-)} = 2 \Delta_\phi = d - 2 = p$ and in this sense the operators $\psi_s^{(+)}$ and $\psi_s^{(-)}$ on the $p$-dimensional defect are like a shadow pair.}
\begin{align}\label{pmfamily}
	\psi_s^{(\pm)}\, \quad \widehat{\Delta}_{s}^{(\pm)}=\Delta_\phi \pm|s|.
\end{align}
We recall that we have set $q=2$ and then have the spins $s \in \mathbb{Z} \text{ or } \mathbb{Z}+\frac{1}{2}$, depending on the choice for the $\mathbb{Z}_2$ monodromy, and $[\psi^{(\pm)}_{s}]^\dagger = \psi^{(\pm)}_{-s}$. As we explain in more detail in appendix \ref{app:greensl}, for $s=0$ the two families merge and there is no degeneracy.

In reflection positive setups the spectrum is further constrained. The scaling dimensions $\widehat \Delta$ of any operator $\widehat \op$ on a $p$-dimensional defect need to obey the standard unitarity condition
\begin{align}\label{genpUB}
	\widehat \Delta &\geq \frac{p}{2}-1 \text{ or } \widehat \Delta = 0, & \text{if} \quad p&>2,\nonumber\\
	\widehat \Delta &\geq 0, & \text{if} \quad p&\leq 2.
\end{align}
If the inequality for $p > 2$ is saturated then the operator is necessarily a free field and its correlators must obey $\square \widehat \op = 0$. If $\widehat \Delta = 0$ for any $p$ then the operator is position-independent, $\partial \widehat \op = 0$. We will also assume cluster decomposition and then, by moving the position-independent operator far away, $\vev{\widehat \op \ldots} = \vev{\widehat \op} \vev{\ldots}$. The operator $\widehat \op$ therefore behaves as a multiple of the identity operator and in particular cannot be charged under transverse rotations.

For our operators $\psi^{(\pm)}_s$ these conditions happen to rule out almost all of the $\psi^{(-)}_s$ since their dimensions are $\frac{p}{2} -s$. In more detail, we first of all observe that the $\psi^{(-)}_s$ modes with $s > 1$ all sit below the unitarity bounds. For non-monodromy defects this leaves the $s = 1$ case but it also happens to always be disallowed: it would be below the unitarity bound for $p = 1$, a charged dimension zero operator for $p = 2$, and a free field for $p > 2$ for which the bulk-defect two-point function does not obey its equation of motion. For monodromy defects the condition $s \leq 1$  leaves the $\psi^{(-)}_{\pm \frac{1}{2}}$ operators and these are allowed for all $p > 1$ but not for $p = 1$ because then they would again correspond to a dimension zero operator.

The analysis of the previous paragraph is summarized in the $q=2$ column of table \ref{tab:unirarytable}. The other columns are obtained by repeating the analysis for higher co-dimensions, the detailed computations for which can be found in appendix \ref{app:freedOPE}.
We note that the aforementioned $s = 0$ degeneracy is lifted if $q > 2$ and in that case the $\psi^{(-)}_0$ mode can become a defect identity operator precisely when $q = p + 2$, and also that free defect fields can never appear because the bulk-defect two-point function never obeys the Laplace equation. Below we will demonstrate that defects are necessarily trivial if none of the non-identity $\psi_s^{(-)}$ operators appears, and this leads directly to the main claim given in the introduction: interesting monodromy defects can exists only for $p \geq 2$ and interesting non-monodromy defects only for $q = 3$ and $p \geq 2$.

\begin{table}[h!]
	\begin{center}
		\begin{tabular}{|c || c|}
			\hline
			${\bf s \in \mathbb Z + \frac{1}{2}}$& ${\bf q=2}$ \\
			\hline
			\hline
			${\bf p=1}$& $\psi_{s}^{(+)}$  \bigstrut\\ 
			\hline
			${\bf p\geq 2}$& $\psi_{s}^{(+)}$ and $\psi_{\pm 1/2 }^{(-)}$ \bigstrut\\
			\hline
		\end{tabular}
		\\\bigskip
		\begin{tabular}{|c||c|c|c|c|c|}
			\hline
			${\bf s \in \mathbb Z}$& ${\bf q=2}$ & ${\bf q= 3}$ & ${\bf q = 4}$& ${\bf q = 5}$ & ${\bf q = 6}$\\
			\hline
			\hline
			${\bf p=1}$ & \multirow{4}{*}{$\psi_{s}^{(+)}$} & $\psi_{s}^{(+)}$ and $\id$ & \multicolumn{3}{|c|}{$\psi_{s}^{(+)}$} \bigstrut\\\cline{1-1}\cline{3-6}
			${\bf p=2}$ & & \multirow{3}{*}{$\psi_{s}^{(+)}$ and $\psi_0^{(-)}$} & $\psi_{s}^{(+)}$ and $\id$ & \multicolumn{2}{|c|}{$\psi_{s}^{(+)}$} \bigstrut\\\cline{1-1}\cline{4-6} 
			${\bf p=3}$ & & &\multirow{2}{*}{$\psi_{s}^{(+)}$} & $\psi_{s}^{(+)}$ and $\id$ & $\psi_{s}^{(+)}$ \bigstrut\\\cline{1-1}\cline{5-6}
			${\bf p=4}$ & & & & $\psi_{s}^{(+)}$ & $\psi_{s}^{(+)}$ and $\id$\bigstrut\\
			\hline
		\end{tabular}
	\end{center}
	\begin{center}
		\caption{\label{tab:unirarytable}Table of unitary defect spectrum in the free theory: for monodromy defects with $q = 2$ and half-integer $s$ on the top, and for general non-monodromy defects on the bottom. The pattern in the bottom table continues outside the shown range of $p$ and $q$. For $q > 2$ the listed operators transform as symmetric traceless $SO(q)$ tensors and then $s$ corresponds to its rank.}
	\end{center}
\end{table}

Before concluding this section, let us comment on the bulk-defect coefficients for the operators $\psi_s^{(\pm)}$. As discussed in \cite{Gaiotto:2013nva} and reviewed in appendix \ref{app:greensl}, in order to reproduce the contact term in the Klein-Gordon equation,
\begin{align}\label{phiphifree0}
	\langle \square\phi(x)\,\phi(x')\rangle=-\frac{4\pi^{\frac{p}{2}+1}}{\Gamma\left(\frac{p}{2}\right)}\,\delta^{p+2}(x-x'),
\end{align}
the coefficients of the $\psi_s^{(+)}$ are necessarily fixed to be
\begin{align}\label{bOPEq2}
	|b_{\phi}^{+,s}|^2 + (p-1) |b_{\phi}^{-,s}|^2 =\frac{(\Delta_\phi)_{|s|}}{{|s|}!}.
\end{align}
Note that any phases in $b_{\phi}^{\pm,s}$ can be absorbed in a phase of the corresponding operators $\psi_s^{(\pm)}$, and therefore we can take the bulk operator expansion coefficients to be real and positive. It follows that any freedom in the bulk-defect expansion coefficients is solely due to the appearance of the `$-$' modes, with only one real parameter introduced for every such mode. Without these modes the two-point function is completely fixed.\footnote{The appearance of the `$-$' mode in the free theory was not considered in \cite{Gaiotto:2013nva}. Note that this mode plays an important role in the free hypermultiplet example of \cite{Bianchi:2019sxz}. Furthermore, as a small generalization of our result we note that a very similar analysis applies to conical metric singularities. In that case the only difference is that the transverse spins $s$ do not have to be half-integers. Such singularities are relevant for the computation of Renyi entropies, see for example appendix C of \cite{Bianchi:2015liz} for a computation in the free scalar theory. It would be interesting to understand the appearance of the `$-$' modes in more detail in this context. We thank Lorenzo Bianchi, Chris Herzog and Marco Meineri for raising this point with us.}

\section{Constraining defect interactions}
\label{sec:constrainingDef}
The goal of this section is to derive constraints on the defect spectrum from analyticity requirements on correlation functions in the presence of defects. The bulk
of this section concerns $q = 2$ defects, but our argument can be extended to higher $q$ with only
small changes; the relevant formulae for generic $q$ are collected in appendix \ref{app:3ptfunctions}. Whenever necessary, we comment about results and differences with respect to the higher co-dimension case. The main characters will be the three-point functions involving the free scalar $\phi$ and one or two defect operators $\widehat{\mathcal{O}}$ and $\widehat{\mathcal{O}}'$:
\begin{align}\label{target}
	\langle \phi(x)\widehat{\mathcal{O}}(\vec{x}')\widehat{\mathcal{O}}' (\vec{x}'')\rangle, \quad
	\langle \phi(x_1)\phi(x_2)\widehat{\mathcal{O}} (\vec{x}'')\rangle.
\end{align}
We will show that the bulk-defect operator expansion of these correlators features unphysical singularities, which can be removed only if very special conditions are met.

Even though our analysis can be carried over to any unitary representation of the parallel Lorentz group, we will restrict ourselves to symmetric and traceless tensors of $SO(p)$. We will contract the Lorentz indices with ``parallel'' polarization vectors $\theta^a$, ($a=1,\dots, p$) on the defect and work with polynomials in $\theta$, see for example \cite{Costa:2011mg} for details. Concretely, for any symmetric and traceless $SO(p)$ tensor of spin $j$ we define
\begin{align}\label{polarvect}
	\widehat{\mathcal{O}}^{(j)}_{s}(\theta,\vec{x})&\equiv \theta^{a_1}\dots \theta^{a_j}\widehat{\mathcal{O}}_{s}^{a_1\dots a_j}(\vec{x}), \quad \theta\gbullet \theta=0,
\end{align}
where $\gbullet$ represents the $SO(p)$-invariant scalar product.

\subsection{Bulk-defect-defect three-point functions}
\label{ss:BDD3pt}

Let us consider first the three-point function of one bulk operator $\phi$ and two defect primaries. For simplicity we take one of them, denoted $\widehat{\mathcal{O}}$, to be an $SO(p)$ scalar, and the second one, denoted $\widehat{{T}}$, to be a symmetric and traceless tensor of parallel spin $j$. Without loss of generality we can place the third operator at infinity and so we investigate:
\begin{align}\label{phiOhOh}
	\langle \phi(\vec{x}_1,z,\bar{z})\widehat{\mathcal{O}}_{s_1}(\vec{x}_2)\widehat{T}{}_{s_2}^{(j)} (\theta,\infty)\rangle.
\end{align}
This correlator is completely determined, via the bulk-defect operator expansion, by the defect three-point functions between $\widehat{T},\widehat{\mathcal{O}}$ and the defect modes of the free scalar $\psi^{(\pf)}_s$ introduced above. These are, in turn, constrained by the defect conformal symmetry to be
\begin{align}\label{defect3ptq2gen}
	\langle \psi_{s}^{(\pf)}(\vec{x}_1)\widehat{\mathcal{O}}_{s_1}(\vec{x}_2)\widehat{T}{}_{s_2}^{(j)} (\theta,\infty)\rangle=\frac{\hat{f}^{(\pf)}_{s\widehat{\mathcal{O}}\widehat{T}}}{|\vec{x}_{12}|^{\widehat{\Delta}^{\pf}_{s}+\widehat{\Delta}_{\widehat{\mathcal{O}}}-\widehat{\Delta}_{\widehat{T}}}}P^{(j)}_\parallel(\hat{x}_{12},\theta).
\end{align}
where we should require that
\begin{align}
	s_1+s_2+s=0.
\end{align}
Note that the dependence on the $SO(p)$ spin is captured by a unique polynomial, homogeneous of degree $j$ in the parallel polarization vector \cite{Costa:2011mg}
\begin{align}\label{parjpoly}
	P^{(j)}_\parallel(\hat{x}_{12},{\theta})\equiv\left(- \hat{x}_{12}\gbullet  {\theta}\right)^j, \quad \hat{x}^a\equiv \frac{x^a}{|\vec{x}|}.
\end{align}
By Bose symmetry the three-point function above cannot depend on the operator ordering\footnote{For line defects the three-point functions generically depend on the ordering of the operators on the line. This will be discussed in section \ref{ss:p1continuation}.} and therefore
\begin{align}\label{BoseSym}
	\hat{f}_{\widehat{\mathcal{O}}\psi\widehat{T}}=(-1)^j \hat{f}_{\psi\widehat{\mathcal{O}}\widehat{T}}.
\end{align}
This implies in particular that only even $j$ are allowed if the first two operators are identical. The complete expression for \eqref{phiOhOh} can be obtained by plugging the bulk-to-defect OPE and resumming the contributions from descendants. After some algebra, which we relegate to appendix \ref{app:3ptfunctions}, the result of this procedure is the  \emph{defect channel expansion} of equation \eqref{phiOhOh}. This expansion takes the form
\begin{align}\label{finalphiOhOh}
	\langle \phi(\vec{x}_1,|z|e^{i\varphi})\widehat{\mathcal{O}}_{s_1}(\vec{x}_2)\widehat{T}{}_{s_2}^{(j)} (\theta,\infty)\rangle=\frac{P^{(j)}_\parallel(\hat{x}_{12},{\theta})}{|z|^{\Delta_{\phi}+\widehat{\Delta}_{\widehat{\mathcal{O}}}-\widehat{\Delta}_{\widehat{T}}}}\sum_{\pf \in\{+,-\}}{b_\phi^{(\pf,s)}}{}\hat{f}^{(\pf)}_{s\widehat{\mathcal{O}}\widehat{T}}
	\,\,{e^{-is\varphi}}{}\mathcal{F}_{\pf,s}^{\widehat{\mathcal{O}}\widehat{{T}}}({\hat{\chi}}),\,\quad s=-s_1-s_2.
\end{align}
The defect blocks in this expression are simple Hypergeometric functions of the cross-ratio\footnote{Since the defect three-point functions \eqref{defect3ptq2gen} do not depend on the transverse angle, the dependence on $\varphi$ in \eqref{finalphiOhOh} enters only via the prefactor $e^{i s \varphi}$ in the bulk-to-defect OPE. It follows that the defect blocks will only depend on $|\vec{x}_{12}|$ and $|z|$, so \eqref{chihat} must be the appropriate cross-ratio.}
\begin{align}\label{chihat}
	\hat{\chi}=\frac{|\vec{x}_{12}|^2}{|z|^2}.
\end{align}
and in appendix \ref{app:3ptfunctions} we show they are given by\footnote{The notation employed in eq. \eqref{OOhOhblocks} is a bit loose, since the defect blocks depend on the quantum numbers of the operators $\widehat{{T}}$ and $\widehat{\mathcal O}$ and not on the operators themselves. May the reader forgive this licentious choice.}
\begin{align}\label{OOhOhblocks}
	\mathcal{F}_{\pf,s}^{\widehat{\mathcal{O}}\widehat{{T}}}({\hat{\chi}})=\hat{\chi}^{\kappa_{\pf\widehat{\mathcal{O}}\widehat{T}}+\frac{j}{2}}\, _2F_1\left(1-\frac{p}{2}-j-\kappa_{\pf\widehat{\mathcal{O}}\widehat{T}},-\kappa_{\pf\widehat{\mathcal{O}}\widehat{T}},1-\frac{p}{2}+\widehat{\Delta}_s^{(\pf)};-\frac{1}{\hat\chi }\right),
\end{align}
where we introduced
\begin{align}\label{kappadef}
	\kappa_{\pf\widehat{\mathcal{O}}\widehat{T}} = -\frac{1}{2} (\widehat{\Delta}_s^{(\pf)}+\widehat{\Delta}_{\widehat{\mathcal{O}}}-\widehat{\Delta}_{\widehat{T}}+j).
\end{align}
Notice that the sum on the r.h.s. of \eqref{finalphiOhOh} contains at most two terms. For $ q > 2$ the prefactors in equation \eqref{finalphiOhOh} change a bit but the functional form of the blocks given in equation \eqref{OOhOhblocks} happens to remain the same, with the index $s$ now denoting the rank of an $SO(q)$ symmetric and traceless tensor. We refer the reader to appendix \ref{app:3ptfunctions} for details.

\subsection{Constraints from analyticity}
\label{ss:BDboot}
In equation \eqref{finalphiOhOh} the $\hat{\chi} \to \infty$ limit corresponds to the bulk-defect operator expansion. If we take the opposite limit $\hat{\chi} \to 0$ we are sending $\vec x_{12} \to 0$ and for finite transverse separation the correlator should be analytic at $\vec{x}_{1}= \vec x_2$. However a generic term in \eqref{finalphiOhOh} is not analytic since:\footnote{When $p = 2$ and $j = 0$ the singularity is actually logarithmic in $\hat\chi$ but the coefficient is essentially the same. For $p = 1$ the non-analyticity of  \eqref{OOhOhopeLim} is due to odd powers of $\sqrt{\chi}\sim |\vec{x}_{12}|$. More details on the continuation to $p=1$ are given in section \ref{ss:p1continuation}.}
\begin{align}\label{OOhOhopeLim}
	P^{(j)}_\parallel(\hat{x}_{12},{\theta})\mathcal{F}_{\pf,s}^{\widehat{\mathcal{O}}\widehat{{T}}}({\hat{\chi}})\underset{\hat\chi\rightarrow 0}{\sim}\frac{\left(-{x}_{12}\gbullet {\theta}\right)^j}{|z|^j}{\hat{\chi}}^{1-j-\frac{p}{2}}\frac{  \Gamma \left(j+\frac{p}{2}-1\right)  \Gamma \left(1-\frac{p}{2}+\widehat{\Delta}_{s}^{(\pf)}\right)}{\Gamma (-\kappa_{\pf\widehat{\mathcal{O}}\widehat{T}} ) \Gamma(j+\widehat{\Delta}_{s}^{(\pf)}+\kappa_{\pf\widehat{\mathcal{O}}\widehat{T}} )}+\dots.
\end{align}
Such unphysical singularities must cancel out from the r.h.s. of \eqref{finalphiOhOh}. This can happen either because of a relation among the OPE coefficients $\hat{f}$ or because the quantum numbers are such that \eqref{OOhOhopeLim} does not hold and the block is actually regular. For all $p$ and $q$ we find that either of the following possible scenarios must be realized.
\begin{enumerate}
	\item[1.] In the first scenario the quantum numbers are such that the `generic' relation \eqref{OOhOhopeLim} is valid. Let us first suppose that both the ``$+$'' and ``$-$'' mode are present, then the cancellation of the unphysical singularities can be enforced by the following relation between the OPE coefficients:\footnote{These special relations, which re-emphasize that one should think of the $\psi^{(\pm)}_s$ as shadow pairs, have appeared already in the context of the long-range Ising model \cite{Paulos:2015jfa,Behan:2017emf,Behan:2017dwr,Behan:2018hfx}. This is not surprising, since the latter has a description in terms of a conformal defect of non-integer co-dimension $q$.}
	\begin{align}\label{exactrela}
		\hat{f}^{(-)}_{s\widehat{\mathcal{O}}\widehat{T}}= -\frac{b_\phi^{(+,s)}}{b_\phi^{(-,s)}}\frac{\Gamma \left(1-\frac{p}{2}+\widehat{\Delta}_{s}^{(+)}\right) \Gamma \left(\frac{j+p-\widehat{\Delta}_{s}^{(+)}+\widehat{\Delta}_{\widehat{\mathcal{O}}}-\widehat{\Delta}_{\widehat{T}}}{2} \right) \Gamma \left(\frac{j+p-\widehat{\Delta}_{s}^{(+)}-\widehat{\Delta}_{\widehat{\mathcal{O}}}+\widehat{\Delta}_{\widehat{T}}}{2} \right)}{\Gamma \left(1+\frac{p}{2}- \widehat{\Delta}_{s}^{(+)} \right) \Gamma \left(\frac{j+\widehat{\Delta}_{s}^{(+)}+\widehat{\Delta}_{\widehat{\mathcal{O}}}-\widehat{\Delta}_{\widehat{T}}}{2} \right) \Gamma \left(\frac{j+\widehat{\Delta}_{s}^{(+)}-\widehat{\Delta}_{\widehat{\mathcal{O}}}+\widehat{\Delta}_{\widehat{T}}}{2} \right)}\hat{f}^{(+)}_{s\widehat{\mathcal{O}}\widehat{T}},
	\end{align}
	where we used the shadow relation $\widehat{\Delta}_s^{(+)}+\widehat{\Delta}_s^{(-)}=p$. On the other hand, if the ``$-$'' mode is absent (or equal to the identity operator for which there is no $\chi \to 0$ singularity), then the coefficient of the corresponding ``$+$'' mode must be zero.
	\item[2.] In the second scenario the scaling dimensions align such that \eqref{OOhOhopeLim} is not valid. This can happen if 
	\begin{itemize}
		\item  $\kappa_{\pf\widehat{\mathcal{O}}\widehat{T}}=n$ with $n \in \mathbb{N}$. In other words,
		\begin{align}\label{multi1}
			\widehat{\Delta}_{\widehat{T}}=\widehat{\Delta}_s^{(\pf)}+\widehat{\Delta}_{\widehat{\mathcal{O}}}+j+2n,\quad n \in \mathbb{N}
		\end{align}
		so the dimension of $\widehat{T}$ equals that of a ``double twist'' combination of $\widehat{\mathcal{O}}$ and $\psi_s^{(\pf)}$.
		\item  $j+\kappa_{\pf\widehat{\mathcal{O}}\widehat{T}}+\widehat{\Delta}_s^{(\pf)}=-n$ with $n \in \mathbb{N}$. In other words,
		\begin{align}\label{multi2}
			\widehat{\Delta}_{\widehat{\mathcal{O}}} = \widehat{\Delta}_s^{(\pf)}+ \widehat{\Delta}_{\widehat{T}}+ j+2n,\quad n \in \mathbb{N}
		\end{align}
		so the dimension of $\widehat{\mathcal{O}}$ equals that of a ``double twist'' combination of $\widehat{T}$ and $\psi_s^{(\pf)}$.
	\end{itemize}
\end{enumerate}

As shown in table \ref{tab:unirarytable}, the ``$-$'' family does not occur in a large class of defects and then the second scenario is the only one that can give non-zero three-point functions. This is the case we will focus on below. It is however interesting to point out that even in the other cases the non-triviality of the correlators is entirely due to the appearance of the single ``$-$'' mode compatible with unitarity listed in table \ref{tab:unirarytable}. Including this mode leads to an interesting variation of the usual bootstrap problem because of the ``shadow'' relations \eqref{exactrela} between OPE coefficients. A first numerical analysis of this type of problem already appeared in the context of the long-range Ising model \cite{Behan:2018hfx} and in \cite{Behan:2020nsf} similar equations were analyzed in the context of boundary conditions for free scalar fields.

\subsection{Reconstructing the bulk}
\label{sec:BBD3pt}

We will now go one step beyond the analysis in the previous subsection and consider three-point functions of the type:
\begin{align}\label{phiphiOscal}
	\langle \phi(x_1)\phi(x_2)\widehat{T}_{s}^{(j)} (\theta,\infty)\rangle.
\end{align}
Note that the only allowed Lorentz representation for $\widehat T$ are symmetric traceless tensors. In a `defect channel' these three-point functions become a sum over the sort of three-point functions that we just considered. Importantly, this sum should be able to reproduce the `bulk channel' OPE which corresponds to bringing the two $\phi$ operators together. We will see that this is indeed the case.\footnote{One might try to go even further and also analyze the three-point function of the bulk field, $\vev{\phi \phi \phi}$, in the presence of the defect. We found that this correlation function does not lead to additional constaints. Note that it automatically vanishes for a monodromy defect.}

\subsubsection{The defect-channel expansion}
Our first goal will be to compute the defect channel blocks for the three-point function \eqref{phiphiOscal}. Our starting point is equation \eqref{defect3ptq2gen} specialized to the case where $\widehat{\mathcal O}$ is one of the $\psi_s^{(\pf)}$, which is:
\begin{align}\label{defect3ptq2}
	\langle \psi_{s_1}^{(\pf_1)}(\vec{x}_1)\psi_{s_2}^{(\pf_2)}(\vec{x}_2)\widehat{T}^{(j)}_{s}(\theta,\infty)\rangle= \frac{\hat{f}^{(\pf_1,\pf_2)}_{s_1s_2 \widehat{T}}}{|\vec{x}_{12}|^{\widehat{\Delta}_{s_1}+\widehat{\Delta}_{s_2}-\widehat{\Delta}_{\widehat{T}}}}P^{(j)}_\parallel(\hat{x}_{12},\theta)\,\delta_{s_1+s_2+s,0}.
\end{align} 
The correlator \eqref{phiphiOscal} can be obtained by acting twice with the bulk-to-defect OPE on the three-point functions \eqref{defect3ptq2} and resumming the contributions from descendants. As we show in appendix \ref{app:3ptfunctions} the result of this computation takes a simple form when we specialize to the kinematical configuration where the two bulk operators lie at the same transverse distance from the defect 
\begin{align}\label{CylConf}
	z_1=|z|e^{i\varphi}, \quad z_2=|z|.
\end{align}
In this configuration, the full three-point function takes the following form 
\begin{align}\label{finalDiagq2}
	\langle \phi(\vec{x}_1,|z|e^{i\varphi})&\phi(\vec{x}_2,|z|)\widehat{T}^{(j)}_{s}(\theta,\infty)\rangle=\nonumber\\
	&\frac{P^{(j)}_\parallel(\hat{x}_{12},{\theta})}{|z|^{2\Delta_{\phi}-\widehat{\Delta}_{\widehat{T}}}}\sum_{\pf_i \in\{+,-\}}\sum_{s_1}{b_\phi^{(\pf_1,s_1)}b_\phi^{(\pf_2,s_2)}}{}\hat{f}^{(\pf_1,\pf_2)}_{s_1s_2 \widehat{T}}
	\,\,{e^{-is_1\varphi}}{}\mathcal{F}_{(\pf_1,s_1),(\pf_2,s_2)}^{\widehat{T}}({\hat{\chi}})\delta_{s_1+s_2+s,0}\,.
\end{align}
The defect blocks in the expression above are computed in appendix \ref{app:3ptfunctions} and read
\begin{align}\label{blocksdef3ptq2}
	&\mathcal{F}_{(\pf_1,s_1),(\pf_2,s_2)}^{\widehat{T}}({\hat{\chi}})=
	{{\hat{\chi}}^{\kappa_{\pf_1,\pf_2}+\frac{j}{2}}}\,\nonumber\\
	& _4F_3\left(\overline{\Delta}_{12}-\hat{h}-\frac{1}{2},\overline{\Delta}_{12}-\hat{h},-\kappa_{\pf_1\pf_2} ,-\kappa_{\pf_1\pf_2} -j-\hat{h};\widehat{\Delta}_{s_1}^{(\pf_1)}-\hat{h},\widehat{\Delta}_{s_2}^{(\pf_2)}-\hat{h},2\overline{\Delta}_{12}-2\hat{h}-1;-\frac{4}{{\hat{\chi}} }\right),\nonumber\\
	&\overline{\Delta}_{12}\,\equiv \frac{1}{2}(\widehat{\Delta}_{s_1}^{(\pf_1)}+\widehat{\Delta}_{s_2}^{(\pf_2)}),\quad 
	\kappa_{\pf_1\pf_2}\,\equiv -\frac{1}{2}(\widehat{\Delta}_{s_1}^{(\pf_1)}+\widehat{\Delta}_{s_2}^{(\pf_2)}-\widehat{\Delta}_{\widehat{T}}+j),\quad
	\hat{h}\,\equiv \frac{p}{2}-1,
\end{align}
where $\hat{\chi}$ is the cross-ratio defined in \eqref{chihat}. Again we should emphasize that the result \eqref{blocksdef3ptq2} holds for all $q$, as we show in appendix \ref{app:3ptfunctions}.

\subsubsection{Consistency with the bulk OPE}
\label{ss:bulktoDefBoot}

Our next goal will be to deduce under which conditions the `defect channel' expansion \eqref{finalDiagq2} is consistent with the `bulk channel' OPE. In order to facilitate this analysis, we integrate both sides of \eqref{finalDiagq2} against $e^{i s' \varphi}$ to obtain
\begin{align}\label{finalDiagProj}
	&\frac{1}{2\pi}\int_0^{2\pi} \mathrm{d} \varphi\,\, e^{i s'\varphi}\langle \phi(\vec{x}_1,|z|e^{i\varphi})\phi(\vec{x}_2,|z|)\widehat{T}^{(j)}_{s}(\theta,\infty)\rangle\nonumber\\
	&=\frac{P^{(j)}_\parallel(\hat{x}_{12},{\theta})}{|z|^{2\Delta_{\phi}-\widehat{\Delta}_{\widehat{T}}}}\sum_{\pf \in\{+,-\}}{b_\phi^{(\pf_1,s')}b_\phi^{(\pf_2,s_2)}}{}\hat{f}^{(\pf_1,\pf_2)}_{s',s_2 \widehat{T}}\,\mathcal{F}_{(\pf_1,s'),(\pf_2,s_2)}^{\widehat{T}}({\hat{\chi}})\,\delta_{s'+s_2+s,0}.
\end{align}
In contrast with equation \eqref{finalDiagq2}, the sum on the r.h.s. of the above expression contains at most three terms. In the higher co-dimension case we can analogously integrate the three-point function against the appropriate spherical harmonics, to isolate each $SO(q)$ irrep.
We will now proceed similarly to what we have done in Section \ref{ss:BDboot}. On the one hand, the bulk self-OPE of the free scalar requires analyticity at $\vec{x}_1 = \vec x_2$ for both the original as well as the integrated correlator \eqref{finalDiagProj}. (Note that the identity operator in the $\phi \times \phi$ OPE does not contribute.) On the other hand, for generic values of its parameters the r.h.s. of \eqref{finalDiagProj} becomes singular in this limit. For generic values of $p$ and $j$ the most singular term is given by:
\begin{align}\label{OOOhopeLim}
	P^{(j)}_\parallel(\hat{x}_{12},{\theta})&\mathcal{F}_{(\pf_1,s_1),(\pf_2,s_2)}^{\widehat{T}}({\hat{\chi}})\underset{\hat\chi\rightarrow 0}{\sim}\frac{\left(-{x}_{12}\gbullet {\theta}\right)^j}{|z|^j}{\hat{\chi}} ^{1-j-\frac{p}{2}}\nonumber\\
	&\frac{\Gamma (\widehat{\Delta}_{s_1}^{(\pf_1)}-\hat{h}) \Gamma (\widehat{\Delta}_{s_2}^{(\pf_2)}-\hat{h}) \Gamma (\hat{h}+j) \Gamma (j+\widehat{\Delta}_{\widehat{T}}-1)}{\Gamma (-\kappa_{\pf_1\pf_2} ) \Gamma (j+\widehat{\Delta}_{s_1}^{(\pf_1)}+\kappa_{\pf_1\pf_2} ) \Gamma (j+\widehat{\Delta}_{s_2}^{(\pf_2)}+\kappa_{\pf_1\pf_2} ) \Gamma \left(\frac{\widehat{\Delta}_{s_1}^{(\pf_1)}+\widehat{\Delta}_{s_2}^{(\pf_2)}+\widehat{\Delta}_{\widehat{T}}+j-p}{2} \right)}+\dots.
\end{align}
but there are also powers of $\hat\chi^{\frac{1}{2}(1-j-p+\widehat{\Delta}_{\widehat{T}})}$ and $\hat\chi^{\frac{1}{2}(2-j-p+\widehat{\Delta}_{\widehat{T}})}$ with equally involved coefficients.

As in the previous subsection, we can find constraints by demanding that such singularities are absent in the full correlator. If both the `$-$' and the `$+$' modes are present then there is an interesting spectrum of constraints to be found on both OPE coefficients and scaling dimensions of $\widehat{T}$ that we will not fully report here. Instead, we focus on the case where only the `$+$' modes are present which is important for the rest of the paper. In that case three-point functions can only be non-zero if the second scenario of section \ref{ss:BDboot} is realized. Furthermore, we should take $\widehat{\mathcal O}$ to also be a defect mode of $\phi$, so $\widehat{\mathcal O} = \psi_{s'}^{(+)}$, and after projecting on a given $s'$ as above we get to apply the `double twist' conditions on $\widehat{\Delta}_{\widehat{T}}$ of that scenario twice. Only the first of the two possibilities listed in that subsection then gives a dimension $\widehat{\Delta}_{\widehat{T}}$ that is above the unitarity bound, and we conclude that:
\begin{align}\label{masterphiphi}
\widehat{\Delta}_{\widehat{T}}=\widehat{\Delta}_{s_1}^{(\pf_1)}+\widehat{\Delta}_{s_2}^{(\pf_2)}+j+2n,&\quad n \in \mathbb{N}.
\end{align}
In short, in the absence of the `$-$' modes the OPE of the $\psi^{(+)}_s$  operators contains only `double twist' operators! Armed with this condition we see that there is no further constraint to be gained from equation \eqref{OOOhopeLim}, since the coefficient of the singular term now vanishes and the $\vev{\phi \phi {\widehat T}}$ correlator is analytic at $\vec x_1 = \vec x_2$.

\section{Triviality of defects of dimension 2 and higher}
\label{sec:theorem}

We have established that for many defects the bulk-defect operator expansion of a free scalar field is constrained to only contain operators that we called $\psi^{(+)}_s$, with fixed coefficients. We have also shown that the non-trivial operators in the OPE of the $\psi^{(+)}_s$ must be of a `double twist' type. In this section we will show that the latter statement implies that all the correlation functions of the $\psi^{(+)}_s$ operators must be those of a generalized free theory. From this the triviality of the $n$-point functions of $\phi$ follows immediately.\footnote{To be precise, when $q = p + 2$ the identity operator can also appear in the bulk-defect operator expansion of $\phi$. Its coefficient is the only variable not fixed by our analysis.}

We will consider here the case $p \geq 2$. The line with $p = 1$ will be discussed in the next section. To apply the theorem below to our analysis of defects with $q > 2$ one should in principle group the operators in representations of the non-abelian transverse rotation algebra. It is however easy to see that this just produces some extra factors that do not change the gist of the argument. 

The result `GFF spectrum implies GFF $n$-point functions' might be interesting on its own; for $n=4$ it can be rephrased as the statement that a trivial double discontinuity \cite{Caron-Huot:2017vep} in all channels implies that the correlation function itself is trivial.\footnote{For four-point functions of identical operators this theorem is a corollary of theorem 1 in \cite{Turiaci_2018}. In all other cases we believe this result is new.}

\begin{theorem} 
	Consider a conformal theory in more than one dimension with a state-operator correspondence and a discrete spectrum such that cluster decomposition is obeyed. Suppose the theory has a set of scalar operators $\y_s(x)$ whose OPEs take the form
	\begin{align}
		\y_{s_1} \times \y_{s_2} = 
		\d_{s_1,s_2} \id + 
		\text{(operators with twist $\widehat{\D}_{s_1} + \widehat{\D}_{s_2} + 2k$, with $k \in \mathbb N$)}
	\end{align}
	Then all the $n$-point correlation functions of $\y_s(x)$ are those of generalized free fields.
	\label{Theorem 1}
\end{theorem}

The main ingredient in our proof will be a dispersion relation in the complex time plane\footnote{A discussion of the analytic structure of conformal correlation functions can found, for example, in \cite{Hartman:2015lfa}. Recent other work on dispersion relations for four-point functions includes \cite{Carmi:2019cub,Bissi:2019kkx}.} for which we will need the commutator $[\y(x),\y(y)]$. For spacelike separation we write the operator product expansion as\footnote{The attentive reader will have noticed a small change of notation: in this section the operators $\y_s$ are taken to be Hermitian. They should be thought of as the real and imaginary part of the $\y_s^{(+)}$ operators of the previous sections.}
\begin{align}
	\y_{s_1}(x) \y_{s_2}(0) = \frac{\d_{s_1,s_2}}{(x^2)^{\widehat{\D}_{s_1}}} \id + \sum_k \lambda\du{12}{k} \frac{x^{\mu_1} \ldots x^{\mu_{\ell_k}}}{(x^2)^{(\widehat{\D}_{s_1}+\widehat{\D}_{s_2}- \widehat{\D}_k+\ell_k)/2}} \op^k_{\mu_1 \ldots \mu_{\ell_k}}(0)
\end{align}
where in the sum over non-trivial operators $k$ we do not distinguish between primaries and descendants. By assumption $\widehat{\D}_k - \ell_k =\widehat{\D}_{s_1} + \widehat{\D}_{s_2} + 2k$, and therefore every term in the sum only yields non-negative integer powers of $x^2$. Passing to the commutator therefore yields
\be
[\y_{s_1}(x), \y_{s_2}(0)] = \text{disc}\left[\frac{\d_{s_i,s_j}}{(x^2)^{\widehat{\D}_{s_i}}} \right] \, \id
\label{Eqn_commutator = discontinuity},
\ee
which is valid as an operator equation as long as the OPE converges. As usual, operator orderings in the commutator must be understood as the Euclidean time orderings and the discontinuity has support only when the operators are causally connected. We will not need the detailed expression of the discontinuity but it is straightforward to work out.\footnote{For integer $\Delta$ the discontinuity is supported only at $x^2 = 0$, in agreement with Huygens' principle.}

\begin{figure}
	\centering 
	\begin{subfigure}{.5\textwidth}
		\centering
		\includegraphics[width=8cm]{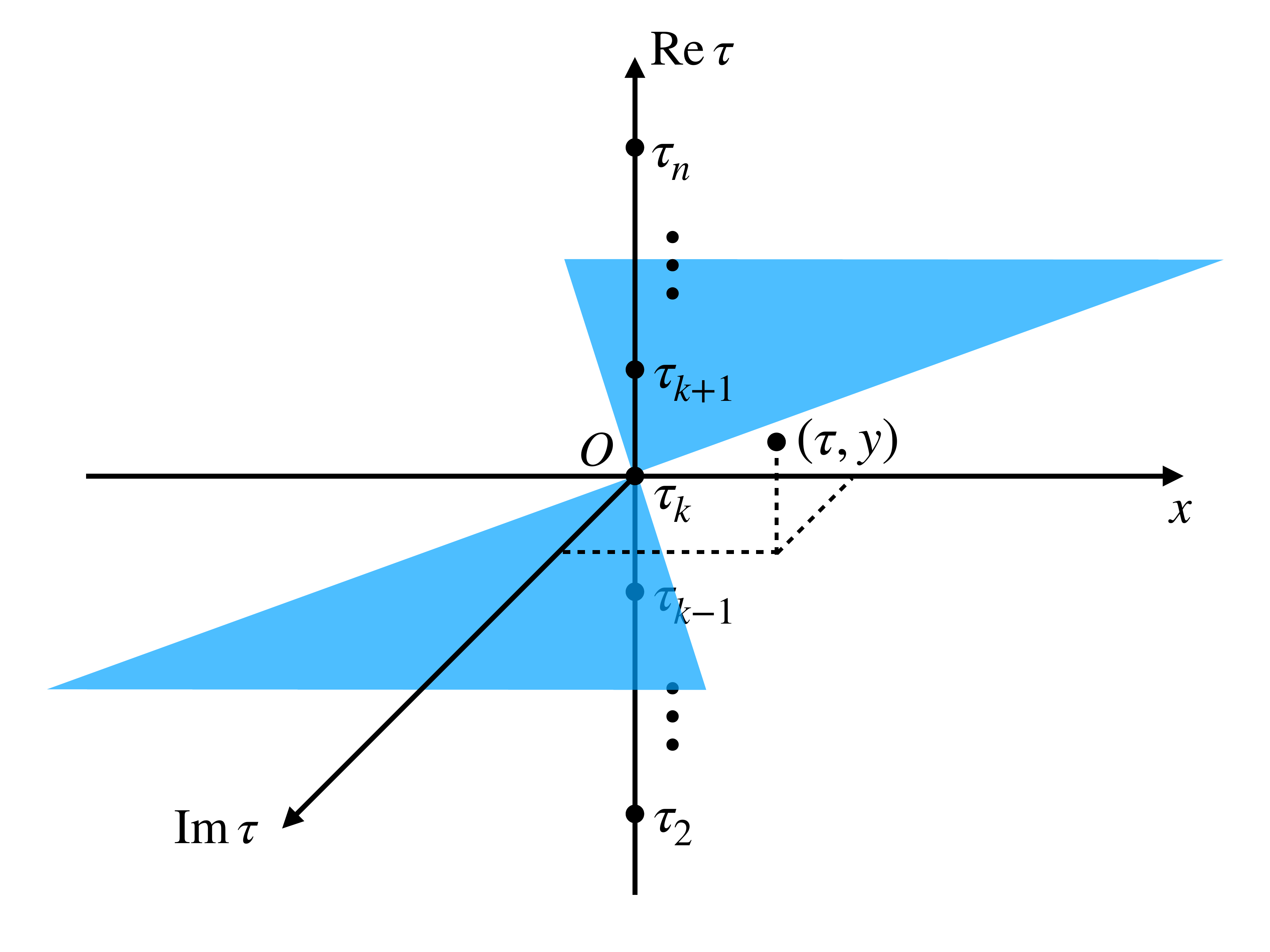}
	\end{subfigure}\hfill%
	\begin{subfigure}{.5\textwidth}
		\centering
		\includegraphics[width=8cm]{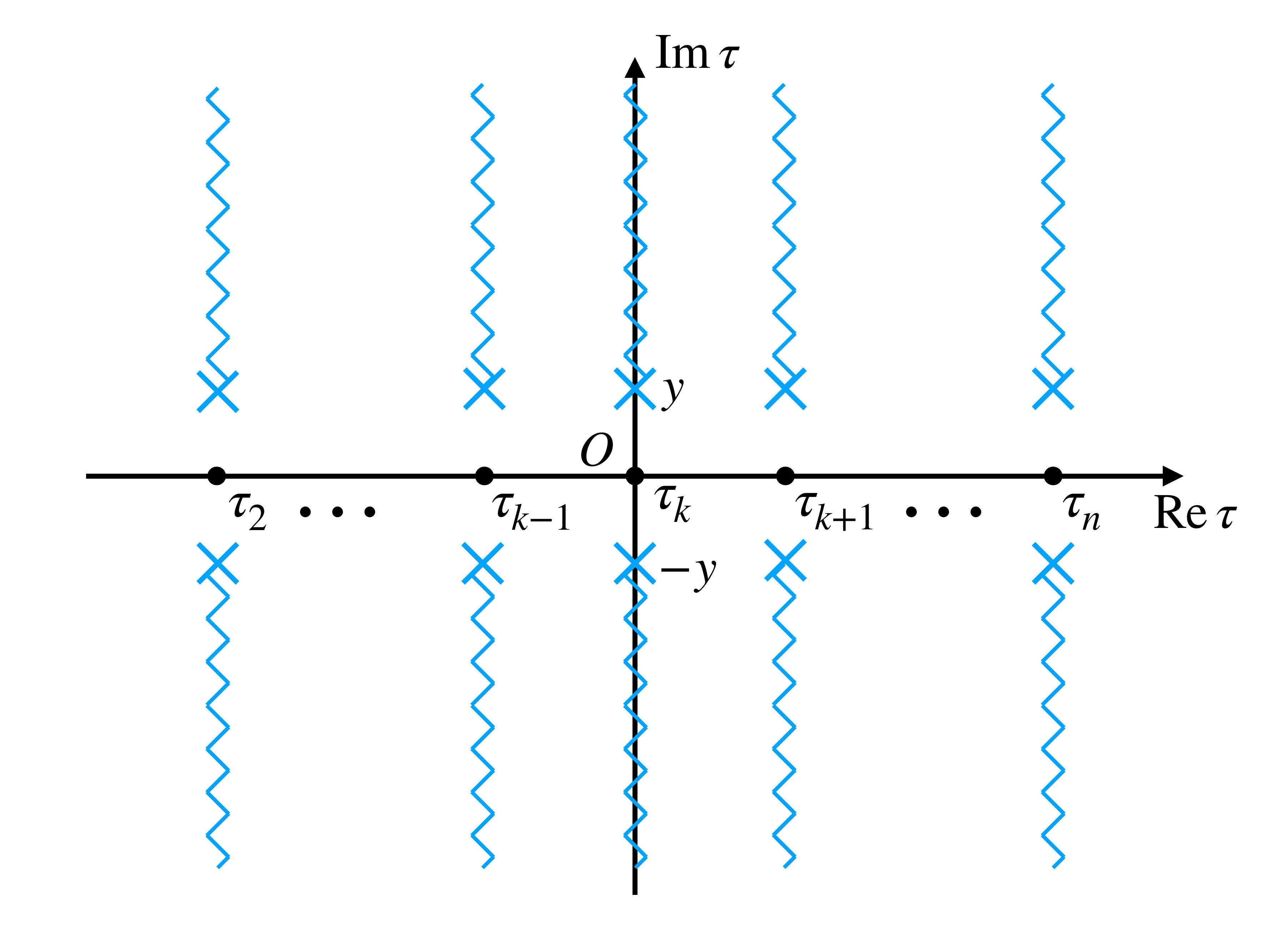}
	\end{subfigure}\hfill%
	\caption{Left: Operators $\y_{s_k}$ $(k\geq2)$ are inserted along the Euclidean time (Re$\,\tau$) axis. Lorentzian time is along the imaginary-$\tau$ axis. The lightcone of $\y_{s_k}$ is illustrated in blue triangles. $\y_{s_1}$ is off the line and its time component $\tau$ is complex in general. Right: Lightcone branch cuts on the complex-$\tau$ plane.}
	\label{Fig_Dispersion setup}
\end{figure}

Our proof will now proceed inductively. We will study the $n$-point function
\be
\vev{\y_{s_1}(x_1) \y_{s_2}(x_2) \y_{s_3}(x_3) \ldots \y_{s_n}(x_n)}
\ee
in the following specific kinematic configuration. We put all operators but the first one on a line:
\be
x_k^\mu = (\tau_k, 0, 0, 0, \ldots ), \qquad\, 2 \leq k \leq n\,,
\ee
ordered such that $\tau_k < \tau_{k+1}$, whereas for the first operator we choose:
\be
x_1^\mu = (\tau, y, 0, 0, \ldots)
\ee
with $y > 0$ a social distancing parameter and $\tau$ an arbitrary \emph{complex} time coordinate. In the $\tau$ plane the correlator is analytic except on the vertical lightcone cuts starting at $\tau = \tau_k \pm i y$ for $2 \leq k \leq n$ and running off to $\pm i \infty$ (see figure \ref{Fig_Dispersion setup}). The discontinuities across these cuts completely determine the correlator because it vanishes at large $|\tau|$ by cluster decomposition (see below for more details). This can be formalized as a dispersion relation:\footnote{Single-variable and two-variable dispersion relations in CFT were recently studied in \cite{Bissi:2019kkx} and \cite{Carmi:2019cub} respectively.}
\begin{align}
	\label{firstdispersion}
	\begin{split}
		G_n(\tau) = \oint \frac{d\tau'}{2\pi i} & \frac{1}{\tau' - \tau} G_n(\tau')\\
		=
		\int_{-\infty}^{\infty} \frac{dt'}{2\pi} 
		&\Big( \frac{1}{\tau - \tau_2 - i t'}  \vev{[\y_{s_1}(\tau_2 + i t, y), \y_{s_2}(\tau_2)] \y_{s_3}(\tau_3) \y_{s_4}(\tau_4)\ldots \y_{s_n}(\tau_n)}\, \\
		&+\frac{1}{\tau - \tau_3 - i t'} \vev{\y_{s_2}(\tau_2)  [\y_{s_1}(\tau_3 + it, y), \y_{s_3}(\tau_3)] \y_{s_4}(\tau_4)\ldots \y_{s_n}(\tau_n)}\, \\
		&+\ldots \, \\
		&+\frac{1}{\tau - \tau_n - i t'} \vev{\y_{s_2}(\tau_2) \y_{s_3}(\tau_3)\ldots  [\y_{s_1}(\tau_n + it, y), \y_{s_n}(\tau_n)]} \Big),
	\end{split}	
\end{align}
where we used that Euclidean time ordering determines the operator ordering.

Next we would like to substitute OPEs and conclude that only the identity contributes in each commutator as in equation \eqref{Eqn_commutator = discontinuity}. Before doing so we need to ensure that the OPE actually converges. Consider the contribution of the commutator between the first and the $k$'th operator in the dispersion relation. It will only have support if the operators become timelike separated, so if $|t'| > y$. OPE convergence for all values of $t'$ can be shown by mapping the configuration into the more familiar $z$ and $\bar z$ coordinates. These can be found by performing a M\"obius transformation:
\be
x \mapsto \frac{x - \tau_k}{x - \tau_{k-1}} \frac{\tau_{k+1} - \tau_{k-1}}{\tau_{k+1} - \tau_k}
\ee
which maps $\tau_k$ to the origin, $\tau_{k+1}$ to 1 and $\tau_{k-1}$ to infinity; the image of operator 1 then defines what we call $z$ and $\bar z$. One finds:
\be
z = \frac{\tau - \tau_k + i y}{\tau - \tau_{k-1} + i y} \frac{\tau_{k+1} - \tau_{k-1}}{\tau_{k+1} - \tau_k} \qquad \qquad \bar z = \frac{\tau - \tau_k - i y}{\tau - \tau_{k-1} - i y} \frac{\tau_{k+1} - \tau_{k-1}}{\tau_{k+1} - \tau_k}
\ee
The M\"obius transformation maps the other operators somewhere on the real axis between $1$ and $\infty$. As is familiar from studies of conformal four-point functions, the desired OPE converges for any $z$ and $\bar z$ away from the real interval $[1, \infty)$, even if we take $z$ and $\bar z$ complex and independent.\footnote{After doing a further transform to the configuration corresponding to the $\rho$ coordinates of \cite{Hogervorst:2013sma} we find the OPE yields an absolutely convergent expansion in powers of $\rho \bar \rho = e^{2 \tau}$ with coefficients that are polynomials in $\sqrt{\rho/\bar \rho} + \sqrt{\bar \rho / \rho} = 2 \cos(\theta)$ with $\tau$ and $\theta$ cylinder coordinates and $\tau < 0$. Adding an imaginary part to $\tau$ clearly does not affect the convergence. Adding an imaginary part to $\theta$ may seem more problematic, but since the twists of all the non-trivial operators are non-negative we can rewrite the expansion as an absolutely convergent series in $\rho$ and $\bar \rho$. This series will then still converge when $\rho$ and $\bar \rho$ are complex and independent, as long as they both have a modulus smaller than one.} Fortunately the entire $t'$ integration region stays in that region: substituting $\tau = \tau_k + i t$ shows that the imaginary parts of $z$ and $\bar z$ are never zero for $|t| > y$. Therefore, the OPE converges and we can substitute equation \eqref{Eqn_commutator = discontinuity} in equation \eqref{firstdispersion}.

It is useful to analyze the large $|\tau|$ behavior in the $z$ and $\bar z$ variables. One finds that $z, \bar z \to (\tau_{k+1} - \tau_{k-1})/(\tau_{k+1} - \tau_k)$ which is a real number greater than one.\footnote{This is actually at the very limit of the domain where the OPE between operator 1 and operator $k$ converges, which illustrates that OPE convergence is not at all guaranteed for a less judicious choice of the $n-1$ operator insertions.} This being the image of infinity, we conclude that there is no operator inserted at this point and the correlator in the $z$, $\bar z$ coordinates does not blow up.\footnote{In fact, the operators at $\tau_{k}$, $\tau_{k+1}$, $\tau_{k+2}$, \ldots $\tau_n$ lie to the left of this point and the other ones to its right, with $\tau_{k-1}$ as stated at infinity. Fusing these two groups of operators together yields a natural OPE channel for this point which converges for large $|\tau|$ both in the Euclidean window and in between the cuts emanating from $\tau_k$ and $\tau_{k+1}$.} It follows that the original correlator must vanish due the Jacobian factor proportional to $(\tau + i y - \tau_{k-1})^{-\Delta} (\tau - i y - \tau_{k-1})^{-\Delta}$. This is the `cluster decomposition' that we alluded to above.

Now let us return to equation \eqref{firstdispersion}. Since each commutator is proportional to the identity operator, each of the $(n-1)$ terms in the dispersion relation factorizes into an $n-2$ point function times the discontinuity of the two-point function, and it is the latter that contains all the $t'$ dependence. The $t'$ integrals are therefore easily done and we find that
\be
\begin{split}
	G_n(t_1) = 
	&\frac{\d_{s_1,s_2}}{(x_{12}^2)^{\widehat{\D}_{s_2}}} \vev{\y_{s_3}(\tau_3) \y_{s_4}(\tau_4)\ldots \y_{s_n}(\tau_n)}\, \\
	&+\frac{\d_{s_1,s_3}}{(x_{13}^2)^{\widehat{\D}_{s_2}}} \vev{\y_{s_2}(\tau_2) \y_{s_4}(\tau_4)\ldots \y_{s_n}(\tau_n)}\, \\
	&+\ldots \,\\
	&+\frac{\d_{s_1,s_n}}{(x_{1n}^2)^{\widehat{\D}_{s_3}}} \vev{\y_{s_2}(\tau_2) \y_{s_3}(\tau_3)\ldots \y_{s_{n-1}}(\tau_{n-1})}.
\end{split}	
\ee
By the induction hypothesis all the $(n-2)$-point functions in the preceding expression are generalized free correlation functions, which when $n$ is even are given by a sum over the $(n-3)!!$ possible complete Wick contractions. For the $n$-point function the above expression gives $(n-1) \times (n-3)!! = (n-1)!!$ terms, and indeed it is easy to verify that this is again just the sum of all possible Wick contractions. We can therefore do induction in steps of two, using the one- and two-point function as a starting point. This also means that the correlation functions vanish for odd $n$, in agreement with the more general result of the previous section.

To complete the proof we need to relax the restricted kinematics where all but one of the operators sit on a straight line. This is straightforward: our argument also goes through for descendants of $\y_s$, so we are free to take any number of derivatives in any given direction acting on any of the $n$ operators. The analyticity of the Euclidean correlation functions away from contact points then dictates that our correlation functions are also equal to those of the generalized free theory for more general choices of the insertion points.

\section{Triviality of line defects}
\label{ss:4ptfunctions}

In this section we consider line defects with $p = 1$. The equivalent of the rotation group on the line is $O(1) \simeq \mathbb Z_2$ which is really just a parity symmetry.\footnote{In the context of the 3d Ising model, this parity has been called S-parity of \cite{Gaiotto:2013nva,Billo:2013jda}.} An important assumption in what follows is that this symmetry is preserved. For definiteness we will take the bulk scalar to be parity even and leave the parity odd case as an exercise.

The main objective of this section is then to prove that there is no room for interacting line defects in our setup, either with or without a monodromy. To this end we will first discuss how the results of section \ref{sec:constrainingDef} can be adapted to the special case of $p=1$. We then adapt the theorem of the previous section and again prove that the ``double twist'' spectrum of defect operators implies that their correlation functions must be those of a generalized free theory.

\subsection{Analytic continuation to line defects}
\label{ss:p1continuation}
Let us first revisit the results of section \ref{sec:constrainingDef}. For the sake of simplicity we will again take $q=2$, but also comment on the main differences with respect to the higher co-dimension case, where necessary. For line defects with parity the only allowed representations for the parallel spin $j$ are the scalar with $j=0$ and the pseudo-scalar with $j=1$. The parity action is given by
\begin{align}
	x\rightarrow \mathcal{R}\,x=-x, \quad \widehat{\mathcal{O}}^{\mathcal{R}(j)}(\mathcal{R}x)=(-1)^j\widehat{\mathcal{O}}^{(j)}(x).
\end{align}
We also recall table \ref{tab:unirarytable}, which states that only the $\psi_s^{(+)}$ modes are allowed in the bulk-to-defect OPE of the free scalar. So to prove the triviality of all line defects it suffices to prove that those modes are generalized free.

The kinematics of correlation functions in the presence of line defects can be obtained from their higher dimensional counterparts presented in section \ref{sec:constrainingDef}. The $O(1)$ spin dependence is captured by the polynomials \eqref{parjpoly} for $j = 0$ or $j=1$. With this in mind, the \emph{most general} three-point function between the defect modes $\psi_{s}^{(+)}$ and any other two defect operators reads
\begin{align}\label{defect3ptq2genLine}
	\langle \psi_{s}^{(+)}({x}_1)\widehat{\mathcal{O}}_{s_1}^{(j_1)}({x}_2)\widehat{T}{}_{s_2}^{(j_2)} (\infty)\rangle=\frac{\hat{f}^{(+)}_{s\widehat{\mathcal{O}}\widehat{T}}}{|{x}_{12}|^{\widehat{\Delta}^{(+)}_{s}+\widehat{\Delta}_{\widehat{\mathcal{O}}}-\widehat{\Delta}_{\widehat{T}}}}(\text{sign}\,x_{12})^{j}\,\delta_{s+s_1+s_2,0},\qquad j\equiv j_1+j_2 \mod 2.
\end{align}
Note that, because of the $\text{sign}$ function above, the defect correlators may depend on the cyclic order of the operators on the line, which is preserved by the conformal algebra but reversed by the parity operation. The operator ordering along the line will play an important role later in this section, when we will be interested in $n$-point correlation functions of the $\psi$'s.

We can now repeat the arguments of sections \ref{ss:BDD3pt} and \ref{ss:BDboot} to obtain constraints on the defect spectrum from the analyticity properties of correlators like
\begin{align}
	\langle \phi({x}_1,z,\bar{z})\widehat{\mathcal{O}}_{s_1}^{(j_1)}({x}_2)\widehat{T}{}_{s_2}^{(j_2)} (\infty)\rangle.
\end{align}
The defect channel expansion of such correlators is again derived by acting with the bulk-defect operator expansion on the three-point functions \eqref{defect3ptq2genLine} and resumming the descendants. It is easy to verify that the result is the natural continuation of \eqref{finalphiOhOh} to $p=1$:
\begin{align}\label{finalphiOhOhLine}
	\langle \phi({x}_1,|z|e^{i\varphi})\widehat{\mathcal{O}}_{s_1}^{(j_1)}({x}_2)\widehat{T}{}_{s_2}^{(j_2)} (\infty)\rangle=\frac{(\text{sign}\,x_{12})^{j}}{|z|^{\Delta_{\phi}+\widehat{\Delta}_{\widehat{\mathcal{O}}}-\widehat{\Delta}_{\widehat{T}}}}{b_\phi^{(+,s)}}{}\hat{f}^{(+)}_{s\widehat{\mathcal{O}}\widehat{T}}
	\,\,{e^{-is\varphi}}{}\mathcal{F}_{+,s}^{\widehat{\mathcal{O}}\widehat{{T}}}({\hat{\chi}}),\,\quad s=-s_1-s_2,
\end{align}
where $j\equiv j_1+j_2 \mod 2$ and with blocks given by \eqref{OOhOhblocks}. Compared to the result obtained earlier for generic $p>1$ -- see equation \eqref{finalphiOhOh} --  the r.h.s. of \eqref{finalphiOhOhLine} contains only a single defect block.\footnote{In the case where $q=3$ the identity operator is also allowed to appear which gives a disconnected contribution.}
Hence, from the analyticity argument of section \ref{ss:BDboot}, we immediately conclude that the defect three-point function \eqref{defect3ptq2genLine} vanishes unless
\begin{align}\label{GFFline}
	\widehat{\Delta}_{\widehat{T}}=\widehat{\Delta}_s^{(+)}+\widehat{\Delta}_{\widehat{\mathcal{O}}}+j+2n,&\quad n \in \mathbb{N}, \quad j\equiv j_1+j_2 \mod 2,\nonumber\\
	&\text{or}\nonumber\\
	\widehat{\Delta}_{\widehat{\mathcal{O}}}=\widehat{\Delta}_s^{(+)}+\widehat{\Delta}_{\widehat{T}}+j+2n,&\quad n \in \mathbb{N}, \quad j\equiv j_1+j_2 \mod 2.
\end{align}
In particular, when $\widehat{\mathcal{O}}$ is itself a defect mode of $\phi$ we find that the scaling dimension of $\widehat{T}_{s}^{(j)}$ must equal
\begin{align}\label{doubletrace}
	\widehat{\Delta}_{\widehat{T}}=\widehat{\Delta}_{s_1}^{(+)}+\widehat{\Delta}_{s_2}^{(+)}+j+2n,&\quad n \in \mathbb{N}.
\end{align}

In conclusion, by repeating the analysis of section \ref{sec:constrainingDef} for line defects, we have proven that the $\psi \times \psi$ OPE contains only operators with ``double twist'' spectrum. In the next section we will argue that the $n$-point functions of the $\psi$'s on the line must again necessarily be those of a generalized free field. 

\subsection{Line defects and generalized free field theories}
In this subsection we will discuss the one-dimensional version of theorem \ref{Theorem 1}. We will again write equations for the co-dimension 2 case, but the generalization to higher co-dimensions is again straightforward.

\begin{theorem}
	Consider a conformal theory in one dimension with parity, a convergent operator product expansion and a discrete spectrum such that cluster decomposition is obeyed. Suppose the theory has scalar operators $\y_s(x)$, with even parity, whose OPEs take the form
	\begin{equation}
		\y_{s_1} \times \y_{s_2} = 
		\d_{s_1,s_2} \id + \text{(operators with twist $\D_{s_1} + \D_{s_2} + 2k$, with $k \in \mathbb N$, and spin $j\in \{0,1\}$)}
	\end{equation}
	Then all the $n$-point correlation functions of the $\y_s(x)$ are those of generalized free fields.
\end{theorem}

The main idea of this proof is similar to that of Theorem \ref{Theorem 1}, that is, we use a dispersion relation and use the GFF spectrum to compute all the discontinuities. However, a subtlety arises in one-dimensional CFT because two correlation functions with different operator ordering modulo cyclic permutations are generically not related by an analytic continuation. For example, if we start with the correlator $\vev{\y_1(0)\y_2(z)\y_3(1) \y_4(\infty)}$ with $0 < z < 1$ and analytically continue it into the complex $z$ plane then its value at negative real $z$ generally does not agree with $\vev{\y_2(z)\y_1(0)\y_3(1) \y_4(\infty)}$. In our case we start with an $n$-point function with operators sequentially ordered,
\begin{align}
	\langle\psi_{s_1}(x_1)\psi_{s_2}(x_2)\ldots\psi_{s_n}(x_n)\rangle, 
	\qquad 
	x_1<x_2<\ldots<x_n,
	\label{n point function physical}
\end{align}
and we would like to explore the complex $x_1$ plane. Suppose we continue $x_1$ via the upper half plane to a real value between $x_2$ and $x_3$. We can use the above OPE to see what happens. If we include the position dependence then it becomes
\begin{align}
	\psi_{s_1}(x_1)\psi_{s_2}(x_2) = \frac{\d_{s_1,s_2}\id}{(x_2-x_1)^{2\widehat{\D}_{s_1}}}
	+\sum_{k} \frac{\lambda\du{12}{k}}{(x_2-x_1)^{\widehat{\D}_{s_1}+\widehat{\D}_{s_2}-\widehat{\D}_k}} \OO^k(x_2),
	\qquad
	x_1<x_2,
\end{align}
By the main assumption of the theorem the contribution of the non-trivial operators gives an analytic contribution in $x_1$ in the vicinity of $x_2$. For the identity operator, on the other hand, we generally obtain a cut and some more detail is needed. We will put the cut in the lower half of the $x_1$ plane, which we emphasize by writing
\be
\psi_{s_1}(x_1)\psi_{s_2}(x_2) = \frac{\d_{s_1,s_2} e^{i \pi \widehat{\D}_{s_1}} \id}{\left( e^{i \pi /2} (x_2-x_1) \right)^{2 \widehat{\D}_{s_1}}}
+\sum_{k} \frac{\lambda\du{12}{k}}{(x_2-x_1)^{\widehat{\D}_{s_1}+\widehat{\D}_{s_2}-\widehat{\D}_k}} \OO^k(x_2)
\ee
and the fractional power is now understood to be evaluated on the principal branch. The analytic continuation via the upper half plane leads to the \emph{analytically continued OPE} given by:
\begin{align}
	\psi_{s_1}(x_1)\psi_{s_2}(x_2) = \frac{\d_{s_1,s_2} e^{i \pi \widehat{\D}_{s_1}} \id}{\left( e^{- i \pi /2} (x_1-x_2) \right)^{2 \widehat{\D}_{s_1}}}
	+\sum_{k} \frac{\lambda\du{12}{k} (-1)^{j_k}}{(x_1-x_2)^{\widehat{\D}_{s_1}+\widehat{\D}_{s_2}-\widehat{\D}_k}} \OO^k(x_2),
	\qquad
	x_2<x_1,
\end{align}
Now we use the assumed parity symmetry. It dictates that
\begin{align}
	\lambda\du{21}{k} =(-1)^{j}\lambda\du{12}{k},
\end{align}
and this allows us to claim that, up to the contribution of the identity operator, the analytically continued OPE is the same as the re-ordered OPE. In equations:
\be
\label{reorderingisok}
\psi_{s_1}(x_1)\psi_{s_2}(x_2) = \frac{\d_{s_1,s_2} (e^{2 i \pi \widehat{\D}_{s_1}} -1)\id}{\left( x_1-x_2 \right)^{2 \widehat{\D}_{s_1}}}
+ \psi_{s_2}(x_2) \psi_{s_1}(x_1), \qquad x_1 > x_2\,.
\ee

\begin{figure}
	\centering
	\includegraphics[width=10cm]{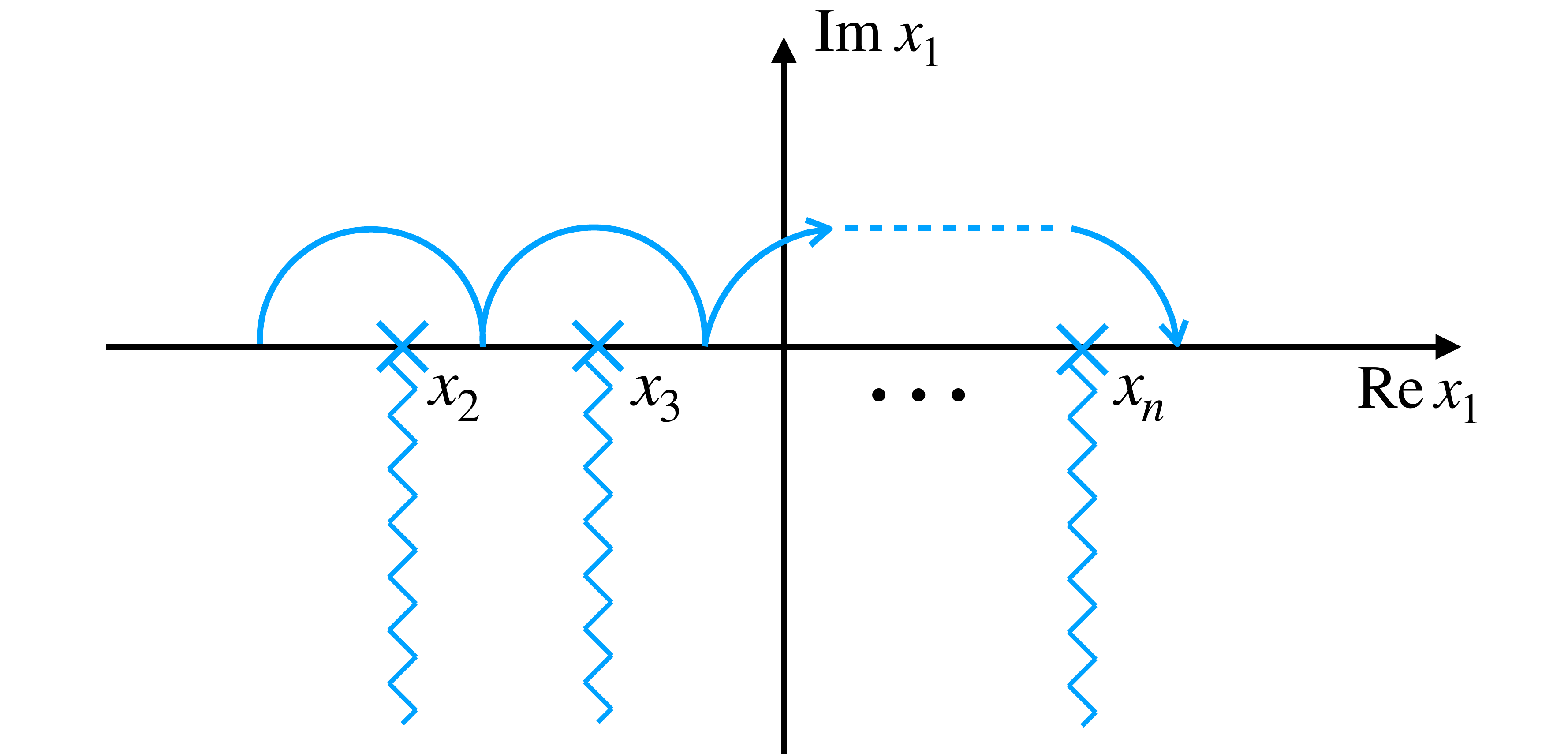}
	\caption{Analyticity structure of $n$-point correlation function is established by consecutively hopping around the $n-1$ operators in the complex $x_1$ plane. The branch cuts are chosen to stretch along the negative imaginary direction.}
	\label{Fig_analyticity structure}
\end{figure}

This is a useful formula. Indeed, since our original correlation function had a $\vev{\psi_1 \psi_2 \psi_3 \ldots}$ ordering, it would normally be impossible to use an OPE to analyze what happens when $x_1$ approaches $x_3$ via analytic continuation. Equation \eqref{reorderingisok} however shows us that, up to a factor proportional to the identity operator, this approach is actually determined by the OPE in the $\vev{\psi_2 \psi_1 \psi_3 \ldots}$ correlator. In terms of the dispersion relation, this indicates that no unwanted contribution arises because the extra factor has vanishing discontinuity along the lightcone branch cut of $\psi_{s_3}, \psi_{s_4},...,\psi_{s_n}$. Of course we can now keep hopping around the remaining $n - 2$ operators and discover the full analytic structure of the $n$-point function in the complex $x_1$ plane: with our choice of cuts, there are vertical branch cuts starting at $x_2, x_3, \ldots x_n$ and no other singularities (see figure \ref{Fig_analyticity structure}). What is more, the discontinuity across those cuts is proportional to a two-point function times an $(n-2)$-point function.\footnote{OPE convergence along the discontinuity is guaranteed by the same arguments as before.} Since the correlation function also falls off\footnote{We claim that any correlation function $\vev{\ldots \psi(x) \ldots}$ in one dimension (that obeys cluster decomposition and without operator insertions at infinity) vanishes when sending $|x| \to \infty$ not along the real axis. To see this, first perform a conformal inversion such that the new variable $x'= -1/x$ goes to zero. We claim that this inverted correlator is finite and the original correlation function therefore vanishes as $|x|^{-2\Delta_\psi}$, which is the Jacobian factor from the inversion. To prove finiteness, suppose first that $\psi(x)$ was the leftmost or rightmost operator. Then finiteness is immediate: the point $x' = 0$ is a physical point with no operators touching. If $\psi(x)$ was not the leftmost or rightmost operator then we need to hop over the other operators to reach $x'=0$. For the correlator at hand the discontinuity is known and does not lead to a further singularity, so finiteness of the inverted correlator follows. But the result is actually more general: just fuse the operators to the left and to the right of $0$ to obtain a sum over three-point functions like $\vev{\psi(x') {\mathcal O}(y_L) {\mathcal O}(y_R)}$ with $y_L < 0$ and $y_R > 0$ the fusion points. The sum converges absolutely in a domain determined by the coordinate differences $|x_i' - y_{L/R}|$ and one can always find a path in the complex $x'$ plane to reach $x'=0$ without exiting this domain. The sum is therefore finite.} at large $|x_1|$ it is once more completely determined by these discontinuities, and the same arguments as those in the previous section show that it equals the generalized free correlation function for all $n$.

\section{Tests in conformal perturbation theory}
\label{sec:confpertthy}

In this section we present some tests of our claims in the context of co-dimension two defects.
In all the examples we consider, the candidate conformal defect is obtained by coupling the free bulk scalar to the local operators of a lower-dimensional CFT$_p$ living on the defect, and flowing to the IR. In the UV the interaction is taken to be
\begin{align}\label{cptansatz}
	S_{\text{int}}= \sum_I g_I \int_{\mathbb{R}^p} \mathrm{d}^p x \, \widehat{\varphi}_I (\vec{x}),
\end{align}
where $\widehat{\varphi}_I$ are some scalar composites made of (derivatives of) the bulk fields as well as of local operators in the CFT$_p$. We then seek for IR fixed points of \eqref{cptansatz} which allow for non-trivial bulk-to-defect interactions. If all the operators $\widehat{\varphi}_I$'s have dimensions $\widehat{\Delta}_I=p-\delta_I$ with $0< \delta_I\ll 1$, then the deformation \eqref{cptansatz} is slightly relevant and the RG flow can be studied perturbatively. At the first order in couplings $g_I$ the beta functions read \cite{cardy}
\begin{align}\label{firstOrderFlow}
	\beta_K= -\delta_K g_K +\frac{S_{p-1}}{2}\sum_{I,J}\widehat{f}_{IJK}\,\,g_I g_J+O(g^3).
\end{align}
In the equation above, $S_{p-1}$ is the volume of the $(p-1)$-sphere and the numbers $\widehat{f}_{IJK}$, which are real in unitary theories, denote the three-point functions of the $\varphi$'s computed at $g_I=0$
\begin{align}
	\widehat{f}_{IJK}\equiv \langle\widehat{\varphi}_I(0)\widehat{\varphi}_J(1)\widehat{\varphi}_K(\infty)\rangle.
\end{align}

A simple example of this scenario is the case where $\varphi_I = \phi$, which is marginal when $q = p + 2$ and was also considered in \cite{Billo:2016cpy}. In agreement with our main result, the only effect of this defect is to give a one-point function to $\phi$ that is proportional to $g$. Below we will present some other concrete realizations of the flow \eqref{cptansatz}. For these examples, we will check explicitly (using conformal perturbation theory) that the CFT$_p$ decouples from the bulk at the unitary IR fixed points, whenever the $\psi^{(-)}$ modes are not generated.

\subsection{Coupling the trivial defect to lower-dimensional matter}

We start by considering a slightly relevant deformation which couples the defect limit of $\phi$ to some operator $\widehat{\mathcal{O}}$ of a given CFT$_p$. If $\widehat{\mathcal{O}}$ has dimension $\frac{p}{2}-\delta$, with $0 \leq \delta \ll 1$, then a natural coupling is
\begin{align}\label{line3dRG}
	S_{\text{int}} = \int_{\mathbb{R}^p} \mathrm{d}^px\,\left(g_1\,\phi (\vec{x},0) \widehat{\mathcal{O}}(\vec{x})+g_2\,\phi^2 (\vec{x},0)\right).
\end{align}
Note that the ``single-trace'' coupling controlled by $g_1$ generates the marginal operator $\phi^2$ at the leading order along the RG flow. The interaction \eqref{line3dRG} has the form of \eqref{cptansatz} with $\widehat{\varphi}_1\equiv\phi\, \widehat{\mathcal{O}}$ and $\widehat{\varphi}_2\equiv\phi^2$. From the general result \eqref{firstOrderFlow} it is straightforward to obtain the beta functions at the first order:
\begin{align}
	\beta_{1} =-\delta g_1 +\frac{S_{p-1}}{2}\widehat{f}_{211}\,g_1 g_2,\quad 
	\beta_{2} =\frac{S_{p-1}}{2}\left(\widehat{f}_{222}\,g_2^2+\widehat{f}_{112}\,g_1^2\right).
\end{align}
From the second equation above, it is clear that a unitary and non-trivial fixed point of this deformation will exist only if $\widehat{f}_{222}$ and $\widehat{f}_{112}$ have opposite sign. On the other hand, these numbers can be computed using Wick's theorem (since the bulk and the defect are decoupled at $g_1=g_2=0$), and as such they are product of two-point functions. Since two-point functions, in turn, must be positive in unitary theories, we conclude that the only possible fixed point at this order is the trivial one, and the CFT$_p$ is decoupled.

\vspace{10pt}
For a concrete realization of the ``single-trace'' deformation \eqref{line3dRG} we can consider the Yukawa coupling of a 4d free scalar field to a 2d free fermion $\chi$:
\begin{align}\label{scalarQED}
	S_{\text{int}} =  \int_{\mathbb{R}^2}\mathrm{d}^2x\,\left( g_1\bar{\chi}\chi\phi+g_2\phi^2+g_3 (\bar{\chi}\chi)^2\right).
\end{align}
Since the fermion has UV dimension $\widehat{\Delta}_\chi=\frac{p-1}{2}=\frac{1}{2}$, the Yukawa coupling is classically marginal. As soon as we turn on $g_1$, the marginal couplings $\phi^2$ and $(\bar{\chi}\chi)^2$ will be generated at one-loop. From \eqref{firstOrderFlow}, the beta functions at the first order read
\begin{align}\label{margcase}
	\beta_{1} =\frac{S_{p-1}}{2}\widehat{f}_{131}\,g_1 g_3,\quad 
	\beta_{2} =\frac{S_{p-1}}{2}\left(\widehat{f}_{222}\,g_2^2+\widehat{f}_{112}\,g_1^2\right),\quad 
	\beta_{3} =\frac{S_{p-1}}{2}\left(\widehat{f}_{333}\,g_3^2+\widehat{f}_{113}\,g_1^2\right).
\end{align}
As in the previous example, the three-point function coefficients above, which can be computed in free theory, are positive numbers. We conclude that the only possible fixed point of \eqref{scalarQED} is the trivial one. 

The story becomes more interesting if we dimensionally continue \eqref{scalarQED} below dimension two while keeping the co-dimension fixed, i.e. work with $p=2-\delta$ and $d=4-\delta$ ($0\leq \delta \ll 1$). When doing so, the operator $\phi^2$ remains marginal, while $g_1,g_3$ become slightly relevant and therefore we find
\begin{align}
	\beta_{1} =-\frac{g_1\delta}{2}&+\frac{S_{p-1}}{2}\widehat{f}_{131}\,g_1 g_3,\quad
	\beta_{2} =\frac{S_{p-1}}{2}\left(\widehat{f}_{222}\,g_2^2+\widehat{f}_{112}\,g_1^2\right),\nonumber\\
	&\beta_{3} =-g_3\delta+\frac{S_{p-1}}{2}\left(\widehat{f}_{333}\,g_3^2+\widehat{f}_{113}\,g_1^2\right).
\end{align}
Assuming unitarity, the first two equations set $g_1=g_2=0$, while from the third we get $g_3\sim {\delta}{}$. In other words, the deformation \eqref{scalarQED} flows towards a decoupled Gross–Neveu model.

\subsection{A nearly marginal deformation in free theory}
As another case of the general setup discussed at the beginning of this section, we can try the classically marginal deformation of the free theory
\begin{align}\label{line3dRGmass}
	S_{\text{int}} = \frac{g}{2}\int_{\mathbb{R}^p} \mathrm{d}^px\,\, \phi^2 (x,0).
\end{align}
Although scale invariance is preserved classically, this example turns out to break it in a subtle way quantum-mechanically. To establish this fact, it is sufficient to compute the exact bulk-to-defect correlator $\langle \phi (x)\phi (0)\rangle_g$. In order to simplify the task, we will work in $p$-dimensional momentum space and consider
\begin{align}
	\langle \phi (\vec{k},|z_1|\phi (-\vec{k},0)\rangle_{g}.
\end{align}
The tree-level contribution can be extracted from the propagator obtained in appendix \ref{app:greensl} and reads
\begin{align}
	G_{\phi\widehat{\phi}}(\vec{k},|z_1|)\equiv\langle \phi (\vec{k},|z_1|)\phi (-\vec{k},0)\rangle_{g=0} = \frac{C_\phi}{(2\pi)}K_0 (|\vec{k}||z_1|).
\end{align}
The propagator between two ${\phi}(\vec{k},0)$'s on the defect, which can be obtained by taking the limit as $z_1\rightarrow 0$ of the expression above, contains a $\log |z_1|$ divergence. Setting $|z_1|=\mu$ the leading term in this expansion is
\begin{align}\label{propmom}
	G_{\widehat{\phi}\widehat{\phi}}(\vec{k})\equiv\langle \phi (\vec{k},0)\phi (-\vec{k},\mu)\rangle_{g=0} = -\frac{C_\phi}{(2\pi)}(\gamma+\log (|\vec{k}|\mu)-\log 2),
\end{align}
up to subleading terms as the scale $\mu$ is sent to zero. Because of this $\log$, the propagator \eqref{propmom} depends on the scale. This dependence is only ``small'', since the $\mu$-derivative of \eqref{propmom} maps exactly to a contact term in position space, and as such it can be understood as a ``small'' conformal anomaly \cite{Petkou:1999fv,vanRees:2011ir} and not something to worry about. On the other hand, order by order in perturbation theory, the corrections to the bulk-to-defect correlator are geometric and they can be exactly resummed:
\begin{align}\label{propmom2}
	\langle \phi (\vec{k},|z_1|)\phi (-\vec{k},0)\rangle_{g}=G_{\phi\widehat{\phi}}(\vec{k},|z_1|)\sum_{n=0}^{\infty}\left(-g\,G_{\widehat{\phi}\widehat{\phi}}(\vec{k},\mu)\right)^n =\frac{G_{\phi\widehat{\phi}}(\vec{k},|z_1|)}{1+g\,  G_{\widehat{\phi}\widehat{\phi}}(\vec{k},\mu)}.
\end{align}
In this final expression, the $\mu$ dependence is far from being a contact term in position space since
\begin{align}
	\mu \frac{\partial}{\partial \mu}\langle \phi (\vec{k},|z_1|)\phi (-\vec{k},0)\rangle_{g} = \frac{C_\phi}{(2\pi)}  \frac{g\,G_{\phi\widehat{\phi}}(\vec{k},|z_1|)}{\left(1+g\,  G_{\widehat{\phi}\widehat{\phi}}(\vec{k},\mu)\right)^2},
\end{align}
and, as such, it cannot be interpreted as a ``small'' conformal anomaly. The ``small'' conformal anomaly of \eqref{propmom} has exponentiated, leading to a ``large'' breaking of scale invariance and we conclude that the deformation \eqref{line3dRGmass} does not lead to a non-trivial conformal defect.

\subsection{A monodromy defect in free theory}
For our final example we couple a free scalar with non-trivial $\mathbb{Z}_2$ monodromy to a lower dimensional CFT$_p$ equipped with an additional $SO(2)_I$ global symmetry. If the latter contains in its spectrum a operator $\widehat{\mathcal{O}}_{s}$ of dimension $\widehat{\Delta}=\frac{p-1}{2}-\delta$, charged under $SO(2)_I$ with spin $|s|=1/2$, then we can consider an interaction that preserves the diagonal of $SO(2)\times SO(2)_I$:
\begin{align}\label{line3dRGmon}
	S_{\text{int}} = g\int \mathrm{d}^px\,\, \psi_{\frac{1}{2}}^{(+)}\widehat{\mathcal{O}}_{-\frac{1}{2}}+\text{c.c}.
\end{align}
This coupling is consistent with unitarity if $p>1$ and it is slightly relevant if $0 < \delta \ll 1$. 

Since the three-point functions of $\psi_{1/2}^{(+)}$ vanish due to $SO(2)$ symmetry, the existence of an IR fixed point for the interaction \eqref{line3dRGmon} depends on certain complicated conditions that arise at the next-to-leading order in conformal perturbation theory.\footnote{At the next-to-leading order the existence of the fixed point depends on the sign of certain regularized integrals of the four-point function of the deformation, see e.g. \cite{Komargodski:2016auf,Behan:2017emf} and also \cite{Behan:2017mwi} for the case of 1d CFTs.} 

Assuming that there exists a non-trivial fixed point $g^2\sim \delta$, one may wonder how this would fit in with our claims of the previous sections. The applicability of our theorem to defect setups hinges on the absence of the so-called $\psi^{(-)}$ modes in the bulk-defect operator expansion of the bulk field $\phi$. For $p=1$ these modes are excluded by cluster decomposition, but for $p > 1$ our theorem would still dictate that the dynamics of the CFT$_p$ must decouple from the bulk if we could consistenly set these modes to zero.

As it happens, a deformation of the form \eqref{line3dRGmon} necessarily induces the appearance of the $\psi^{(-)}$ modes in the bulk-to-defect OPE of $\phi$ at order $g$. To establish this, it is sufficient to note that the bulk-to-defect two-point function between $\phi$ and $\widehat{\mathcal{O}}_{-1/2}$ is non-zero at order $g$:
\begin{align}
	\langle \phi(0,z,\bar{z})\widehat{\mathcal{O}}_{-s}(0)\rangle =& -g \int \mathrm{d}^p w\,\,\frac{\bar{z}^{s}}{(|z|^2+|\vec{w}|^2)^{\Delta_\phi+s}}\frac{C_{\widehat{\mathcal{O}}\widehat{\mathcal{O}}}}{ |\vec{w}|^{2\Delta_\phi-2s-2\delta}}+O({g^2})\nonumber\\
	=&-g\, C_{\widehat{\mathcal{O}}\widehat{\mathcal{O}}} S_{p-1} \frac{\bar{z}^{s}}{|z|^{p}}\frac{\Gamma \left(\frac{p}{2}\right) \Gamma (s)}{2 \Gamma \left(\frac{p}{2}+s\right)}+O(\delta^{\frac{3}{2}}), \quad s=1/2.
\end{align}
This result matches the expected form of a correlator between $\phi$, and $\psi_{-1/2}^{(-)}$ -- see equation \eqref{phiO2pt} -- with bulk-to-defect coefficients $b_\phi^{-,-1/2}\propto g$.

In conclusion, the deformation \eqref{line3dRGmon} could provide an example of a unitary, non-trivial conformal defect for $p > 1$. The way it is allowed to be non-trivial is consistent with our theorem.

\section{Applications}
\label{sec:discussion}
For the single free scalar field $\phi$ the space of possible conformal defects is remarkably constrained, and in many cases the only allowed defects are trivial in the sense specified in the introduction. In all dimensions and co-dimensions the appearance of a $\psi^{(-)}$ mode in the bulk-to-defect expansion of $\phi$ is a necessary condition for a defect to not be trivial.

We should point out that our results also apply when the bulk theory has a decoupled real free scalar, since one can integrate out all the other bulk matter and conclude that the $n$-point functions of the scalar are trivial. This works in particular for some supersymmetric theories. For example, consider the surface defects in the abelian $(2,0)$ theories in six dimensions that were recently discussed in \cite{Drukker:2020dcz}. If conformality is not spoiled by an anomaly as in the example discussed above, we would expect triviality of the scalar correlation functions. Another example  are the aforementioned defects in the $\mathcal N = 2$ four-dimensional free hypermultiplet. These appear to be labelled by a monodromy $\Phi \to e^{i \alpha} \Phi$ with $\Phi$ a complex scalar. We can immediately conclude that no defect can exist for $\alpha = 0$ and that for other values of $\alpha$ any non-triviality is allowed because of the single $\psi^{(-)}$ mode in the two-point function of the bulk scalar. This is in line with some recent explicit computations in \cite{Bianchi:2019sxz}. In three dimensions our results also match with the supersymmetric literature: for example, the non-trivial defects in free $\mathcal N= 4$ theories in \cite{Dimofte:2019zzj} all appear to have a scale associated with them.

Defects and free scalar fields also naturally appear as the infrared description of vortices and Goldstone bosons in setup where a $U(1)$ global symmetry is spontaneously broken. In that case the size $R$ of the vortex provides a natural scale. Our results imply that interactions between the Goldstone degrees of freedom and the vortex trivialize in the deep infrared when $R \to 0$. A good physical example of this situation is the scattering of phonons off a vortex in superfluid helium. In this case a microscopic model is available, and computations in for example \cite{Kozik_2005} (but see also references therein) confirm this view.

Let us finally point out an interesting possible extension of our results to conformal defects in (weakly) interacting theories. As explained in section \ref{sec:constrainingDef} there are always unphysical singularities when we apply the bulk-defect operator expansion to three-point functions, and their cancellation will therefore imply a infinite and non-trivial sum rule. It would be interesting to analyze tese constraints further, for example in an epsilon expansion or in a large $N$ limit.

\section*{Acknowledgements}
We thank the organizers of the `Bootstrap 2019' and the `Boundaries and Defects in Quantum Field Theory' workshops at Perimeter Institute where part of this work was carried out. We thank L.~Bianchi, D.~Gaiotto, C.~Herzog, M.~Lemos and M.~Meineri for pointing out imperfections in the first version of this paper.  We are grateful to C. Behan and L.~Di~Pietro for collaboration on related topics. We further thank S.~Cremonesi, D.~Gaiotto, A.~Gimenez-Grau, M.~Lemos, D.~Maz\'a\v c and V.~Schomerus for interesting discussions and especially Miguel Paulos for numerous illuminating comments. EL and BvR thank the Perimeter Institute for Theoretical Physics for hospitality. Research at Perimeter Institute is supported by the Government of Canada through Industry Canada and by the Province of Ontario through the Ministry of Research \& Innovation.
EL, BvR and XZ are supported by the Simons Foundation grant $\#$488659 (Simons Collaboration on the non-perturbative bootstrap).
PL is supported by the DFG through the Emmy Noether research group ``The Conformal Bootstrap Program'' project number 400570283.  

\appendix

\section{Details of the scalar bulk-to-defect OPE}
\label{app:freedOPE}

In this appendix we give some details about the bulk-to-defect OPE of a scalar bulk operator. We then specialize to the case where the bulk operator is a free field, and we spell out the constraints imposed by the equations of motion on the spectrum of its defect modes.

For the sake of completeness, we consider generic conformal defects of dimension $p$ and co-dimension $q$. In order to encode the $SO(p)\times SO(q)$ spin it is convenient to contract the corresponding indices with ``parallel'' or ``transverse'' polarizations vectors, respectively $\theta^a$ $(a=1,\dots,p)$ and $w^i$ $(i=1,\dots,q)$, and work with polynomials in these variables. The following definitions generalizes the ones given in \eqref{polarvect}
\begin{align}\label{polarvectgen}
	\widehat{\mathcal{O}}^{a_1\dots a_j}_{s}(w,\vec{x})&\equiv w^{i_1}\dots w^{i_s}\widehat{\mathcal{O}}^{a_1\dots a_j}{}_{i_1\dots i_s}(\vec{x}), \quad w\wbullet w=0,\nonumber\\
	\widehat{\mathcal{O}}^{(j)}_{s}(w,\theta,\vec{x})&\equiv \theta^{a_1}\dots \theta^{a_j}\widehat{\mathcal{O}}_{s}^{a_1\dots a_j}(w,\vec{x}), \quad \theta\gbullet \theta=0,
\end{align}
where the symbols $\wbullet$ and $\gbullet$ represent, respectively, $SO(q)$-invariant and $SO(p)$-invariant scalar products in real space.

The bulk-to-defect OPE of a scalar bulk operator $\Sigma(x)$ contains infinitely many defect primaries $\widehat{\Sigma}_s$, scalars under $SO(p)$ and transforming as symmetric and traceless tensors of $SO(q)$. If we neglect the contribution from defect descendants, we have schematically
\begin{align}\label{bOPEO}
	\Sigma(x)=\sum_{\widehat{\Sigma},s} \frac{b_{\Sigma}^{\widehat{\Sigma}}}{|x_\perp|^{\Delta_{\Sigma} -\widehat{\Delta}_{\widehat{\Sigma}}}}\,(w\wbullet \hat{x})^s\,\widehat{\Sigma}_s(w,\vec{x})+\dots
\end{align}
In the expression above we introduced the unit vector $\hat{x}^i\equiv \frac{x^i}{|x_\perp|}$, orthogonal w.r.t. the defect.\footnote{To recover the operator's contribution in real space from the expression above it is sufficient to note that $\,(w\wbullet \hat{x})^s\,\widehat{\Sigma}_s(w,\vec{x})$ is mapped to $\,\hat{x}^{i_1}\dots \hat{x}^{i_s}\,\widehat{\Sigma}^{i_1\dots i_s}(\vec{x})$.} If we take the defect operators to be unit normalized, then the numbers $b_{\Sigma}^{\widehat{\Sigma}}$ are identified precisely with the bulk-to-defect couplings
\begin{align}\label{BDcouplingsgen}
	\langle \Sigma(x) \widehat{\Sigma}_s(w,0)\rangle = \frac{b_{\Sigma}^{\widehat{\Sigma}}(w\wbullet \hat{x})^s}{|x_\perp|^{\Delta_{\Sigma}-\widehat{\Delta}_{\widehat{\Sigma}}}(x^2)^{\widehat{\Delta}_{\widehat{\Sigma}}}}.
\end{align}
The contribution from the defect descendants in \eqref{bOPEO} is completely encoded into \eqref{BDcouplingsgen}. By comparing the bulk-to-defect OPE with \eqref{BDcouplingsgen} one finds \cite{Billo:2016cpy}
\begin{align}\label{bOPEphi}
	\Sigma(x)=\sum_{\widehat{\Sigma},s}\sum_n \frac{b_{\Sigma}^{\widehat{\Sigma}}}{|x_\perp|^{\Delta_{\Sigma} -\widehat{\Delta}_{\widehat{\Sigma}}}}\,\frac{\left(-\frac{1}{4}|x_\perp|^2 \vec \nabla^2_\parallel\right)^n }{n! \left(\widehat{\Delta}_{\widehat{\Sigma}}+1-\frac{p}{2}\right)_n}\,(w\wbullet \hat{x})^s\,\widehat{\Sigma}_s(w,\vec{x}).
\end{align}
We now specialize to the case where the bulk operator $\Sigma$ is a free scalar, which we denote by $\phi$, with defect modes $\psi_s$. As we have shown explicitly for co-dimension two defects in Section \ref{ss:twopoint}, the defect primaries that can couple to $\phi$ are selected by the free equation of motion. Requiring that the Laplacian annihilates \eqref{BDcouplingsgen} at separated points gives the following condition \cite{Billo:2016cpy}
\begin{align}\label{freeTq}
	(\hat{\tau}-\Delta_{\phi})(\Delta_{\phi}-\hat{\tau}+2-q-2s)=0,\qquad \hat{\tau}=\widehat{\Delta}_{\widehat{\Sigma}}-s.
\end{align}
Assuming no further degeneracy, for each spin $s$ at most two families of defect primaries are allowed in the bulk-to-defect OPE of $\phi$. These two solutions, denoted as $\psi_{s}^{(\pm)}$, form a shadow pair on the defect
\begin{align}\label{defSpec}
	\psi_{s}^{(+)}:\quad \widehat{\Delta}_s^{(+)}=\Delta_{\phi}+s,\quad \text{or}\quad 
	\psi^{(-)}_{s}:\quad \widehat{\Delta}_{s}^{(-)}=\Delta_{\phi}+2 -q-s.
\end{align}
Crucially, the spin of the second family is restricted by unitarity \eqref{genpUB} to the values $s\leq \frac{4-q}{2}$ (for $p>1$) and $s\leq \frac{3-q}{2}$ (for $p=1$). Note that for $q= p + 2$ there is a $`-'$ mode of dimension zero and (as explained in the main text) by cluster decomposition we may assume it proportional to the defect identity $\id$. For $p>2$ the primary that saturates the unitarity bound is a free field and must obey the Laplace equation, which is inconsistent with a non-zero two-point function with the bulk field $\phi$. It can therefore be consistently removed from the spectrum. Altogether the unitary defect spectrum is summarized in table \ref{tab:unirarytable} in the main text.

\section{Two-point function in free theory for $q=2$ defects}
\label{app:greensl}

In this appendix we perform the computation of the two-point function of a free scalar in the presence of a twist defect. This computation was originally performed by the authors of \cite{Gaiotto:2013nva}, however we will obtain a slightly more general result. 
The starting point is Green's equation \eqref{phiphifree0}, which we report here for convenience
\begin{align}\label{phiphifree}
	-\square G(x_1,x_2)=C_{\phi}\,\delta^{p+2}(x_1-x_2),\,\,\,C_{\phi}\equiv\frac{4\pi^{\frac{p}{2}+1}}{\Gamma\left(\frac{p}{2}\right)}.
\end{align}
The normalization $C_\phi$ is chosen in such a way that 
\begin{align}
	G(x_1,x_2)\underset{x_1\rightarrow x_2}{\sim}\frac{1}{|x_1-x_2|^{d-2}}.
\end{align}
To solve Green's equation, it is convenient to Fourier transform to the $p$-dimensional momentum space along the defect and then adopt a basis of $SO(2)$ spherical harmonics. In terms of the complex coordinates $z_1=|z_1|e^{i \varphi}$ and $z_2=|z_2|$ we obtain
\begin{align}
	G(x_1,x_2)=\sum_{s} \int \frac{d^p k}{(2\pi)^p}\,\,e^{i \vec{k}\cdot \vec{x}_{12}}e^{i s\varphi}a_s(|\vec{k}|,|z_{1}|,|z_{2}|),
\end{align}
where the sum runs over all (half)-integers, depending on the choice of monodromy for $\phi$. Denoting $|z_i|=r_i$ for simplicity of notation, we find that the modes $a_s$ satisfy the following differential equation \footnote{We used $\delta^2(x-x')=\frac{1}{r}\delta(\varphi-\varphi')\delta(r-r')$ and then $\delta(\varphi)=\frac{1}{2\pi}\sum_s e^{i s \varphi}$.}
\begin{align}\label{homopolarpro}
	\left(|\vec{k}|^2-\frac{\partial^2}{\partial r_1^2}-\frac{1}{r_1}\frac{\partial}{\partial r_1}+\frac{s^2}{r_1^2}\right)a_s(|\vec{k}|,r_1,r_2)=\frac{C_{\phi}}{2\pi} \frac{1}{r_1}\delta(r_1-r_2).
\end{align}
The homogeneous problem has a general solution given by $a_s(|\vec{k}|,r_1,r_2)=A(r_2) I_{|s|} (|\vec{k}| r_1)+B(r_2) K_{|s|}  (|\vec{k}| r_1)$, where $I_s(x),K_s(x)$ are modified Bessel functions. Let us consider the region where $r_1\geq r_2$. Then, regularity of the solution asymptotically far away from the defect, i.e. $r_1\rightarrow \infty$, sets $A=0$. In the region where $r_1\leq r_2$, the $I_{|s|}(|\vec{k}|r)$ are regular while $K_{|s|}$ behave as 
\begin{align}\label{Ksol}
	K_{|s|}  (|\vec{k}|r)\underset{r\rightarrow 0}{\sim}&  |\vec{k}|^{-|s|} r^{-|s|} + |\vec{k}|^{|s|} r^{|s|},\nonumber\\
	K_{0}  (|\vec{k}|r)\underset{r\rightarrow 0}{\sim}&  c\,\log (r |\vec{k}|)+c',
\end{align}
for some constants $c,c'$. Due to the logarithmic singularity in the second line of the above, which is not allowed by conformal invariance, we are forced to set $B=0$ for $s=0$. For $|s|>0$, on the other hand, there is no reason to impose regularity conditions at $r=0$, since we do not expect the physics to be smooth in the proximity of the defect. In terms of the bulk-to-defect OPE, the singular modes in the first line of \eqref{Ksol} take into account the presence of the defect primaries of dimensions $\Delta_\phi-|s|$. These singular solutions are compatible with unitarity as long as $p=1$ or  $p>1$ and $0<|s|< 1$. Hence
\begin{align}\label{homosol}
	a_s(|\vec{k}|,r_1,r_2)=
	\begin{cases}|s|\geq 0,\quad B_s^{\text{II}}(r_2) K_{|s|} (|\vec{k}| r_1),&\quad r_1\geq r_2, \\ 
		|s|=0 \text{ or } |s|\geq 1,\quad A_s^{\text{I}}(r_2) I_{|s|} (|\vec{k}| r_1),&\quad r_1\leq r_2,\\
		0<|s|< 1,\quad A_s^{\text{I}}(r_2) I_{|s|} (|\vec{k}| r_1)+B_s^{\text{I}}(r_2) K_{|s|} (|\vec{k}| r_1),&\quad r_1\leq r_2.\\ 
	\end{cases}
\end{align}
For $p=1$ the solution \eqref{Ksol} is either a constant mode ($|s|=\frac{1}{2}$) or below the unitarity bound and we are free to set it to zero by choosing $B=0$ for all $s$.\footnote{Upon invoking cluster-decomposition principle, as we did in appendix \ref{app:freedOPE}.} Let us now go back to the inhomogeneous problem and fix the solution \eqref{homosol} in order to reproduce the contact term in the r.h.s. of \eqref{homopolarpro}. To this end, we need to impose continuity of \eqref{homosol} at $r_1=r_2$, and that the discontinuity of its first derivative at $r_1=r_2$ equals precisely $\frac{C_\phi}{2\pi r_2}$. 
After some little algebra we find
\begin{align}\label{homosol2}
	a_s(|\vec{k}|,r_1,r_2)=
	\begin{cases}
		B_s^{\text{I}}(r_2)K_{|s|} (|\vec{k}| r_1)+\frac{C_\phi}{2\pi}I_{|s|} (|\vec{k}| r_2)K_{|s|} (|\vec{k}| r_1),&\quad r_1\geq r_2 \\ 
		B_{s}^{\text{I}}(r_2)K_{|s|} (|\vec{k}| r_1)+\frac{C_\phi}{2\pi}I_{|s|} (|\vec{k}| r_1) K_{|s|} (|\vec{k}| r_2),&\quad r_1\leq r_2 \\ 
	\end{cases}
\end{align}
with the understanding that $B_{s}^{\text{I}}(r_2)\neq 0$ only for $p>1$ and $0<|s|<1$. If we finally impose symmetry under exchange of the two external scalars, which are identical, we find the condition
\begin{align}
	B_s^{\text{I}}(r_1)K_{|s|} (|\vec{k}| r_2)=B_s^{\text{I}}(r_2)K_{|s|} (|\vec{k}| r_1),
\end{align}
which is satisfied by $B_s^{\text{I}}(r_2)= h_s K_{|s|} (|\vec{k}| r_2)$, for any real constant $h_s$. The final solution can be written as
\begin{align}\label{solfingen}
	&G(x_1,x_2)=G^{(-)}(x_1,x_2)+G^{(+)}(x_1,x_2)\nonumber\\
	&G^{(-)}(x_1,x_2)=\sum_{s=\pm \frac{1}{2}} \int \frac{d^p k}{(2\pi)^p}\,\,e^{i \vec{k}\cdot \vec{x}_{12}}e^{i s\varphi} h_sK_{|s|} (|\vec{k}| r_{1})K_{|s|} (|\vec{k}| r_{2})\nonumber\\
	&G^{(+)}(x_1,x_2)=\frac{C_\phi}{2\pi}\sum_{s\neq\pm \frac{1}{2}} \int \frac{d^p k}{(2\pi)^p}\,\,e^{i \vec{k}\cdot \vec{x}_{12}}e^{i s\varphi} I_{|s|} (|\vec{k}| r_{<})K_{|s|} (|\vec{k}| r_{>}),
\end{align}
where $r_<=\text{min}(r_1,r_2)$ and $r_>=\text{max}(r_1,r_2)$. Note that the expression above differs from the result of \cite{Gaiotto:2013nva} by the additional contribution $G^{(-)}(x_1,x_2)$.  One can explicitly perform the momentum-space integration in the first line of \eqref{solfingen} to find
\begin{align}
	G^{(-)}(x_1,x_2)=\frac {2^{p-2}h_{\frac{1}{2}}}{(2 \pi )^{\Delta_\phi-\frac{1}{2}}}\,\Gamma\left(\Delta_\phi-\frac{1}{2}\right)\,\frac{\cos\left(\frac{\varphi}{2}\right)}{(r_1 r_2)^{\Delta_\phi}}\frac{\left(\xi + \sqrt{\xi(\xi +4)}+2\right)^{\Delta_\phi-\frac{1}{2}}}{ \left(\xi +\sqrt{\xi(\xi +4)}+4\right)^{2\Delta_\phi-1}},
\end{align}
where we introduced the cross ratio
\begin{align}
	\xi=\frac{|\vec{x}_{12}|^2+(r_1-r_2)^2}{r_1 r_2}.
\end{align}
In real space, the spin $s$ contribution to $G^{(+)}(x_1,x_2)$ is \cite{Gaiotto:2013nva}
\begin{align}\label{Gplus}
	G_s^{(+)}(x_1,x_2)=\frac{\Gamma(\widehat{\Delta}_{s}^{(+)})}{\Gamma\left(\Delta_\phi\right)\Gamma\left(\widehat{\Delta}_{s}^{(+)}-\Delta_\phi+1\right)}\frac{e^{i s\varphi}\xi^{-\widehat{\Delta}_{s}^{(+)}}}{(r_1 r_2)^{\Delta_\phi}}\, _2F_1\left({\widehat{\Delta}_{s}^{(+)}}{},\widehat{\Delta}_{s}^{(+)}-\frac{p-1}{2};2\widehat{\Delta}_{s}^{(+)}-p+1;-\frac{4}{\xi}\right),
\end{align}
where $\widehat{\Delta}_{s}^{(+)}=\Delta_{\phi}+|s|$. Note that the result \eqref{Gplus} is equivalent to \eqref{bulktodefplus} in virtue of the following identity
\begin{align}\label{4F3identity}
	\xi ^{-x} \, _2F_1\left(x,-\frac{p}{2}+x+\frac{1}{2};-p+2 x+1;-\frac{4}{\xi }\right)=(\xi+2) ^{-x} \, _2F_1\left(\frac{x+1}{2},\frac{x}{2};-\frac{p}{2}+x+1;\frac{4}{(\xi+2) ^2}\right).
\end{align}
Note that if we impose trivial monodromy, the result \eqref{Gplus} leads to the two-point function for a trivial defect:
\begin{align}
	\frac{1}{(x_1-x_2)^{d-2}}=\sum_{s \in \mathbb{Z}}G_s^{(+)}(x_1,x_2).
\end{align}
For the twist defect in $p=2$, we note that the generic solution \eqref{solfingen} takes a simple form
\begin{align}
	G(x_1,x_2)=\frac{2\,\cos \left(\frac{\varphi }{2}\right)}{(x_1-x_2)^2}\left(\frac{1 }{\sqrt{\xi +4}}+\frac{h_{\frac{1}{2}}(1-\cos\varphi+\xi/2)}{\sqrt{2}}\frac{   \sqrt{2+\xi +\sqrt{\xi  (\xi +4)}}}{\left(4+\xi +\sqrt{\xi  (\xi +4)}\right)}\right),
\end{align}
which reduces to the result of \cite{Gaiotto:2013nva} when we set $h_{\frac{1}{2}}=0$. 

Finally, by comparing \eqref{solfingen} with the defect channel blocks \eqref{bulktodefplus} we can extract the relevant bulk-to-defect OPE coefficients:
\begin{align}\label{newbulkOPE}
	&|b_{\phi}^{s,+}|^2+(p-1)|b_{\phi}^{s,-}|^2=\frac{(\Delta_\phi)_{|s|}}{{|s|}!},\nonumber\\
	&|b_{\phi}^{s,-}|^2=\delta_{|s|,\frac{1}{2}}\,\frac{h_{\frac{1}{2}}}{4\pi ^{\Delta_\phi-\frac{1}{2}}} \Gamma \left(\Delta_\phi -\frac{1}{2}\right),\nonumber\\
	&0\leq h_{\frac{1}{2}} \leq \frac{4 \pi ^{p/2-1}}{\Gamma \left(\frac{p}{2}\right)}.
\end{align}
The inequality in the last line follows from $|b_{\phi}^{s,\pm}|\geq 0$, which is required by reflection-positivity.

\section{Three-point functions from the bulk-to-defect OPE}
\label{app:3ptfunctions}
This appendix contains the derivations of the defect conformal blocks presented in section \ref{sec:constrainingDef}. In what follows we will keep $p,q$ generic, for the sake of completeness. As a further generalization, we will take the bulk scalar to be generic, i.e. not necessarily free, and denote it as $\Sigma$ (as we did in appendix \ref{app:freedOPE}).

Let us start from deriving the defect expansion of 
\begin{align}\label{3ptfunctionbu}
	\langle \Sigma(x_1)\widehat{\mathcal{O}}_{s_2}(w_2,\vec{x}_2)\widehat{T}{}_{s_3}^{(j)} (w_3,\theta,\infty)\rangle.
\end{align}
In the expression above, $\widehat{\mathcal{O}}$ and $\widehat{T}$ are symmetric and traceless tensors of $SO(q)$, respectively of spin $s_2$ and $s_3$. The dependence on the $SO(p)\times SO(q)$ is encoded into polynomials in the polarization vectors $\{w_i\},\theta$, as explained in appendix \ref{app:freedOPE}. 

The starting point is the three-point functions between the defect modes of $\Sigma$, denoted as $\widehat{\Sigma}$, and any other two defect operators:
\begin{align}\label{OOhOhgen}
	\langle \widehat{\Sigma}_{s_1}(w_1,\vec{x}_1)\widehat{\mathcal{O}}_{s_2}(w_2,\vec{x}_2)\widehat{T}^{(j)}_{s_3} (w_3,\theta, \infty)\rangle=\frac{\hat{f}_{\widehat{\Sigma}\widehat{\mathcal{O}}\widehat{T}}}{|\vec{x}_{12}|^{\widehat{\Delta}_{\widehat{\Sigma}}+\widehat{\Delta}_{\widehat{\mathcal{O}}}-\widehat{\Delta}_{\widehat{T}}}}P^{(s_1,s_2,s_3)}_\perp(\{w_i \})P^{(j)}_\parallel(\hat{x}_{12},\theta).
\end{align}
The $SO(p)$ spin is encoded in the polynomials $P^{(j)}_\parallel$, which were already introduced in \eqref{parjpoly}. The $SO(q)$ spin dependence is captured by the polynomials $P^{(s_1,s_2,s_3)}_\perp$, which are homogeneous of degree $s_i$ in the transverse polarization vectors $w_i$
\begin{equation}\label{SpinPstruc}
	P^{(s_1,s_2,s_3)}_\perp(\{w_i \})\equiv(w_1\wbullet w_2)^{\frac{1}{2} (s_1+s_2-s_3)} (w_1\wbullet w_3)^{\frac{1}{2} (s_1-s_2+s_3)} (w_2\wbullet w_3)^{\frac{1}{2} (s_2-s_1+s_3)},
\end{equation}
where $s_i$ are non-negative integers satisfying
\begin{align}\label{spinCond0}
	s_1+s_2-s_3=2n_1,\quad s_1-s_2+s_3=2n_2,\quad s_2-s_1+s_3=2n_3,\qquad n_i \in \mathbb{N}. 
\end{align}
Note that the $w_1,w_2,w_3$'s cannot be linearly independent for $q=2$, and as such the basis \eqref{SpinPstruc} becomes over-complete. 

To compute \eqref{3ptfunctionbu}, we apply the bulk-to-defect OPE \eqref{bOPEphi} on the three-point functions \eqref{OOhOhgen}. The derivatives in the parallel directions commute with the $SO(q)$ polynomials and, making use of the identity
\begin{align}\label{parallelLapl}
	\nabla_{\vec{x}_{12}}^{2n}\left(\frac{(-\vec{x}_{12}\gbullet \theta)^j}{|\vec{x}_{12}|^{2t}}\right)=4^n (t)_n \left(1+t-j-\frac{p}{2}\right)_n \frac{(-\vec{x}_{12}\gbullet \theta)^j}{|\vec{x}_{12}|^{2t+2n}},
\end{align}
we can find the following series representation
\begin{align}\label{auxsum1}
	\langle \Sigma(x_1)\widehat{\mathcal{O}}_{s_2}(w_2,\vec{x}_2)&\widehat{T}{}_{s_3}^{(j)} (w_3,\theta,\infty)\rangle=\nonumber\\
	&\frac{P^{(j)}_\parallel(\hat{x}_{12},{\theta})}{|x_{1\perp}|^{\Delta_{\Sigma}+\widehat{\Delta}_{\widehat{\mathcal{O}}}-\widehat{\Delta}_{\widehat{T}}}}\sum_{\widehat{\Sigma},s}{b_\Sigma^{\widehat{\Sigma}}}\,\,\hat{f}_{\widehat{\Sigma}\widehat{\mathcal{O}}\widehat{T}}
	\,\,{P_{\perp}^{(s,s_2,s_3)}(\{w_i\})}{}(w\wbullet \hat{x}_1)^s\times\nonumber\\
	&\times{\hat\chi}^{\kappa_{\widehat{\Sigma}\widehat{\mathcal{O}}\widehat{T}}+\frac{j}{2}}\sum_n\frac{({-\hat\chi})^{-n}}{n!}\frac{(-\kappa_{\widehat{\Sigma}\widehat{\mathcal{O}}\widehat{T}} )_n \left(1-\frac{p}{2}-j-\kappa_{\widehat{\Sigma}\widehat{\mathcal{O}}\widehat{T}}\right)_n}{  \left(\widehat{\Delta}_{\widehat{\Sigma}}-\frac{p}{2}+1\right)_n},
\end{align}
where we introduced the parameter
\begin{align}
	\kappa_{\widehat{\Sigma}\widehat{\mathcal{O}}\widehat{T}} = -\frac{1}{2} (\widehat{\Delta}_{\widehat{\Sigma}}+\widehat{\Delta}_{\widehat{\mathcal{O}}}-\widehat{\Delta}_{\widehat{T}}+j),
\end{align}
as well as the cross-ratio \eqref{chihat}. The sum over $s$ is truncated to those values that satisfy $SO(q)$ selection rules \eqref{spinCond0}. Finally, the sum over $n$ can be performed for generic values of the parameters. The results is a beautiful Hypergeometric function
\begin{align}\label{OOhOhblocksgen}
	\mathcal{F}_{\widehat{\Sigma}}^{\widehat{\mathcal{O}}\widehat{{T}}}({\hat{\chi}})=\hat{\chi}^{\kappa_{\widehat{\Sigma}\widehat{\mathcal{O}}\widehat{T}}+\frac{j}{2}}\, _2F_1\left(1-\frac{p}{2}-j-\kappa_{\widehat{\Sigma}\widehat{\mathcal{O}}\widehat{T}},-\kappa_{\widehat{\Sigma}\widehat{\mathcal{O}}\widehat{T}},1-\frac{p}{2}+\widehat{\Delta}_{\widehat{\Sigma}};-\frac{1}{\hat\chi }\right).
\end{align}
When we take the bulk operator to be a free scalar, the expression above gives precisely eq. \eqref{OOhOhblocks}. This result, can also be obtained by solving the relevant Casimir equation as done in \cite{Karch:2018uft}.\footnote{We do not find perfect agreement with the block calculated in \cite{Karch:2018uft}. We obtained the same Casimir equation however our block is a different linear combination of solutions. The solution of \cite{Karch:2018uft} does not seem to be consistent with the OPE limit.}  
Importantly, the defect blocks $\mathcal{F}$ are completely blind to the transverse directions. In particular they only depend on the parallel dimension $p$, and \emph{not} on $q$.

\vspace{10pt}
Let us now consider the bulk-bulk-defect three-point function
\begin{align}\label{3ptfunctionbu2}
	\langle \Sigma(x_1)\Sigma(x_2)\widehat{T}{}_{s}^{(j)} (w,\theta,\infty)\rangle.
\end{align}
Again, we will not require $\Sigma$ to be a free scalar. The complete form of the expression above can be obtained by applying once again the bulk-to-defect OPE to eq. \eqref{auxsum1} and then resum the descendants. In practise it is easier to start from the three-point functions
\begin{align}
	\langle \widehat{\Sigma}_{s_1}(w_1,\vec{x}_1)\widehat{\Sigma}'_{s_2}(w_2,\vec{x}_2)\widehat{T}^{(j)}_{s} (w,\theta, \infty)\rangle=\frac{\hat{f}_{\widehat{\Sigma}\widehat{\Sigma}'\widehat{T}}}{|\vec{x}_{12}|^{\widehat{\Delta}_{\widehat{\Sigma}}+\widehat{\Delta}_{\widehat{\Sigma}'}-\widehat{\Delta}_{\widehat{T}}}}P^{(s_1,s_2,s)}_\perp(\{w_i \})P^{(j)}_\parallel(\hat{x}_{12},\theta),
\end{align}
and apply twice on it the bulk-to-defect OPE \eqref{bOPEphi}. Making use twice of the identity \eqref{parallelLapl} we obtain
\begin{align}\label{summand}
	\langle \Sigma(x_1)\Sigma(x_2)\widehat{T}^{(j)}_{s} (w,\theta,\infty)\rangle=&\sum_{\widehat{\Sigma},\widehat{\Sigma}',s_1,s_2}\,b_{\Sigma}^{\widehat{\Sigma}}\, b_{\Sigma}^{\widehat{\Sigma}'}\, \hat{f}_{\widehat{\Sigma}\widehat{\Sigma}' \widehat{T}}\nonumber\\
	&\times \sum_{m,n}\frac{(-1)^{m+n}}{m! n!}\frac{  {|x_{1\perp}|}^{{\widehat{\Delta}_{\widehat{\Sigma}}-\Delta_\Sigma }{}+2n} {|x_{2\perp}|}^{{\widehat{\Delta}_{\widehat{\Sigma}'}-\Delta_\Sigma }{}+2m}   }{ |\vec{x}_{12}|^{-2\kappa_{\widehat{\Sigma}\widehat{\Sigma}'}+2m+ 2n-j} 
	}\nonumber\\
	& \times \frac{(-\kappa_{\widehat{\Sigma}\widehat{\Sigma}'})_m (-\kappa_{\widehat{\Sigma}\widehat{\Sigma}'}+m)_n \left(-\kappa_{\widehat{\Sigma}\widehat{\Sigma}'}-\hat{h}-j\right)_m \left(-\kappa_{\widehat{\Sigma}\widehat{\Sigma}'}+m-\hat{h}-j\right)_n}{\left(\widehat{\Delta}_{\widehat{\Sigma}}-\hat{h}\right)_n\left(\widehat{\Delta}_{\widehat{\Sigma}'}-\hat{h}\right)_m }\nonumber\\
	&\times\,\underbrace{(w_1\wbullet \hat{x}_1)^{s_1}(w_2\wbullet \hat{x}_2)^{s_2}P^{(s_1,s_2,s)}_\perp(w_1,w_2,w)}_{{\mathcal{W}_{\perp}}^{(s_1,s_2,s)}(\hat{x}_1,\hat{x}_2,w)}P^{(j)}_\parallel(\hat{x}_{12},\theta),
\end{align}
where we introduced
\begin{align}
	\kappa_{\widehat{\Sigma}\widehat{\Sigma}'}\,\equiv- \frac{1}{2}(\widehat{\Delta}_{\widehat{\Sigma}}+\widehat{\Delta}_{\widehat{\Sigma}'}-\widehat{\Delta}_{\widehat{T}}+j),\quad \hat{h}\,\equiv \frac{p}{2}-1.
\end{align}
The integers $s_1,s_2,s$ are constrained by the selection rules \eqref{spinCond0}. Resumming this expression is expected to be hard since this configuration is characterized by three cross-ratios (compare to \eqref{chideftot}),
\begin{align}\label{chidef}
	{{\chi}}\equiv\frac{|\vec{x}_{12}|^2+|x_{1\perp}|^2+|x_{2\perp}|^2}{{|x_{1\perp}|}{|x_{2\perp}|}},\quad t\equiv\frac{{|x_{1\perp}|}}{{|x_{2\perp}|}},\quad  \cos\varphi\equiv\hat{x}_1 \wbullet \hat{x}_2.
\end{align} 
For our purposes, which is studying the bulk OPE limit of \eqref{3ptfunctionbu2}, it will be sufficient to specialize \eqref{summand} to the ``cylindrical'' configuration 
\begin{align}\label{diagconf}
	x_1^i=|z|n_1^i, \quad x_2^i=|z|n_2^i, \quad n_1 \wbullet n_2=\cos\varphi, \quad n \wbullet n=1, \quad t=1,
\end{align}
where the resummation can be performed easily. In terms of the cross-ratio $\hat{\chi}$ defined in \eqref{chihat} we find:
\begin{align}\label{OOOhblocksgen}
	&\mathcal{F}_{\widehat{\Sigma}\widehat{\Sigma}'}^{\widehat{T}}({\hat{\chi}})=
	{{\hat{\chi}}^{-\frac{1}{2}(\widehat{\Delta}_{\widehat{\Sigma}}+\widehat{\Delta}_{\widehat{\Sigma}'}-\widehat{\Delta}_{\widehat{T}})}}\,\nonumber\\
	& _4F_3\left(\overline{\Delta}_{12}-\hat{h}-\frac{1}{2},\overline{\Delta}_{12}-\hat{h},-\kappa_{\widehat{\Sigma}\widehat{\Sigma}'} ,-\kappa_{\widehat{\Sigma}\widehat{\Sigma}'} -j-\hat{h};\widehat{\Delta}_{\widehat{\Sigma}}-\hat{h},\widehat{\Delta}_{\widehat{\Sigma}'}-\hat{h},2\overline{\Delta}_{12}-2\hat{h}-1;-\frac{4}{{\hat{\chi}} }\right),
\end{align}
where we defined $\overline{\Delta}_{12}\,\equiv \frac{1}{2}(\widehat{\Delta}_{\widehat{\Sigma}}+\widehat{\Delta}_{\widehat{\Sigma}'})$. When we take $\Sigma$ to be a free scalar we find precisely the blocks  \eqref{blocksdef3ptq2}. Furthermore, from this result we can recover the blocks for the two-point function shown in eq.\eqref{bulktodefplus} by simply setting $\widehat{\Sigma}=\widehat{\Sigma}'$ and the third operator to be the identity. In this case, the functions ${\mathcal{W}_{\perp}}^{(s_1,s_1,0)}$ become Gegenbauer polynomials of $\cos\varphi$
\begin{align}\label{2pt0spin}
	{\mathcal{W}_{\perp}}^{(s,s)}(\hat{x}_1,\hat{x}_2)=&\left({w_1\wbullet \hat{x}_1}{}\right)^{s} \left({w_2\wbullet \hat{x}_2}{}\right)^{s}(w_1\wbullet w_2)^{s}=\frac{s!}{2^s (\frac{q}{2}-1)_s} C_s^{(\frac{q}{2}-1)}(\cos\varphi).
\end{align}
Finally, after the hypergeometric transformation \eqref{4F3identity}, the ${}_4 F_3$ in \eqref{OOOhblocksgen} simply reduces to \eqref{bulktodefplus}.

\bibliography{bib}
\bibliographystyle{JHEP}

\end{document}